\documentclass[reprint,amsmath,amssymb,pra]{revtex4-2}
\usepackage{graphicx}% Include figure files
\usepackage{dcolumn}% Align table columns on decimal point
\usepackage{bm}% bold math
\usepackage{capt-of}
\usepackage{wrapfig}

\usepackage{lipsum} 

\usepackage{blindtext}
\usepackage{graphicx, caption}
\usepackage{multirow}
\usepackage{float}
\usepackage{subfig}
\usepackage{geometry}
\usepackage{amsmath,amssymb}
\usepackage{amsthm}
\usepackage{bbold}
\usepackage{mathtools}
\usepackage{braket}
\usepackage{booktabs}
\usepackage{hyperref}
\hypersetup{colorlinks=true,linkcolor=blue}

\usepackage{stackengine}
\usepackage{soul}

\usepackage{listings}
\usepackage{tikz}
\usepackage{scrextend}
\usepackage{bm}
\usetikzlibrary{arrows}
\usetikzlibrary{arrows.meta}
\usetikzlibrary{decorations.pathmorphing}
\usetikzlibrary{hobby}
\usetikzlibrary{patterns}
\usetikzlibrary{calc}
\usetikzlibrary{plotmarks}

\newtheorem*{thm*}{Theorem}

\newtheorem*{prop*}{Proposition}

\newtheorem*{lemma*}{Lemma}

\newtheorem*{cor*}{Corollary}

\newtheorem*{cj*}{Conjecture}

\newtheorem*{Def*}{Definition}

\makeatletter
\def\thmhead@plain#1#2#3{%
  \thmname{#1}\thmnumber{\@ifnotempty{#1}{ }\@upn{#2}}%
  \thmnote{ {\the\thm@notefont#3}}}
\let\thmhead\thmhead@plain
\makeatother

\theoremstyle{definition}

\newcommand{\bb}{\begin{equation}\begin{aligned}\hspace{0pt}}
\newcommand{\bbb}{\begin{equation*}\begin{aligned}}
\newcommand{\ee}{\end{aligned}\end{equation}}
\newcommand{\eee}{\end{aligned}\end{equation*}}
\newcommand*{\coloneqq}{\mathrel{\vcenter{\baselineskip0.5ex \lineskiplimit0pt \hbox{\scriptsize.}\hbox{\scriptsize.}}} =}

\DeclareMathAlphabet{\pazocal}{OMS}{zplm}{m}{n}

\newcommand{\lsmatrix}{\left(\begin{smallmatrix}}
\newcommand{\rsmatrix}{\end{smallmatrix}\right)}

\stackMath

\stackMath

\makeatletter
\newcommand*\rel@kern[1]{\kern#1\dimexpr\macc@kerna}
\newcommand*\widebar[1]{%
  \begingroup
  \def\mathaccent##1##2{%
    \rel@kern{0.8}%
    \overline{\rel@kern{-0.8}\macc@nucleus\rel@kern{0.2}}%
    \rel@kern{-0.2}%
  }%
  \macc@depth\@ne
  \let\math@bgroup\@empty \let\math@egroup\macc@set@skewchar
  \mathsurround\z@ \frozen@everymath{\mathgroup\macc@group\relax}%
  \macc@set@skewchar\relax
  \let\mathaccentV\macc@nested@a
  \macc@nested@a\relax111{#1}%
  \endgroup
}

\counterwithin*{equation}{part}
\counterwithin*{thm}{part}
\counterwithin*{figure}{part}

\tikzset{meter/.append style={draw, inner sep=10, rectangle, font=\vphantom{A}, minimum width=30, line width=.8, path picture={\draw[black] ([shift={(.1,.3)}]path picture bounding box.south west) to[bend left=50] ([shift={(-.1,.3)}]path picture bounding box.south east);\draw[black,-latex] ([shift={(0,.1)}]path picture bounding box.south) -- ([shift={(.3,-.1)}]path picture bounding box.north);}}}
\tikzset{roundnode/.append style={circle, draw=black, fill=gray!20, thick, minimum size=10mm}}
\tikzset{squarenode/.style={rectangle, draw=black, fill=none, thick, minimum size=10mm}}

\definecolor{Blues5seq1}{RGB}{239,243,255}
\definecolor{Blues5seq2}{RGB}{189,215,231}
\definecolor{Blues5seq3}{RGB}{107,174,214}
\definecolor{Blues5seq4}{RGB}{49,130,189}
\definecolor{Blues5seq5}{RGB}{8,81,156}

\definecolor{Greens5seq1}{RGB}{237,248,233}
\definecolor{Greens5seq2}{RGB}{186,228,179}
\definecolor{Greens5seq3}{RGB}{116,196,118}
\definecolor{Greens5seq4}{RGB}{49,163,84}
\definecolor{Greens5seq5}{RGB}{0,109,44}

\definecolor{Reds5seq1}{RGB}{254,229,217}
\definecolor{Reds5seq2}{RGB}{252,174,145}
\definecolor{Reds5seq3}{RGB}{251,106,74}
\definecolor{Reds5seq4}{RGB}{222,45,38}
\definecolor{Reds5seq5}{RGB}{165,15,21}

\usepackage{pgfplots}

\begin{document}
\title{Statistical Characterization of Entanglement Degradation Under Markovian Noise in Composite Quantum Systems}
\author{Nunzia Cerrato}
\affiliation{Scuola Normale Superiore, Piazza dei Cavalieri 7, I-56126 Pisa, Italy}
\author{Sauro Succi}
\affiliation{Center for Life Nano- \& Neuro-Science, Italian Institute of Technology (IIT),
viale Regina Elena 295, Rome, 00161, Italy and
Department of Physics, Harvard University, 17 Oxford St, Cambridge, MA 02138, United States}
\author{Giacomo De Palma}
\affiliation{Department of Mathematics, University of Bologna, 40126 Bologna, Italy} 
\author{Vittorio Giovannetti}
\affiliation{NEST, Scuola Normale Superiore and Istituto Nanoscienze-CNR, Piazza dei Cavalieri 7, I-56126 Pisa, Italy}

\begin{abstract}		
Understanding how noise degrades entanglement is crucial for the development of reliable quantum technologies. While the Markovian approximation simplifies the analysis of noise, it remains computationally demanding, particularly for high-dimensional systems like quantum memories. In this paper, we present a statistical approach to study the impact of different noise models on entanglement in composite quantum systems. By comparing global and local noise scenarios, we quantify entanglement degradation using the Positive Partial Transpose Time (PPTT) metric, which measures how long entanglement persists under noise. 
When the sampling of different noise scenarios is performed under controlled and homogeneous conditions, our analysis reveals that systems subjected to global noise tend to exhibit longer PPTTs, whereas those influenced by independent local noise models display the shortest entanglement persistence. To carry out this analysis, we employ a computational method proposed by Cao and Lu, which accelerates the simulation of PPTT distributions and enables efficient analysis of systems with dimensions up to $D=8$. Our results demonstrate the effectiveness of this approach for investigating the resilience of quantum systems under Markovian noise.
\end{abstract}

\maketitle

\section{Introduction}
\begin{figure}[t]
    \centering
    \includegraphics[width=\linewidth]{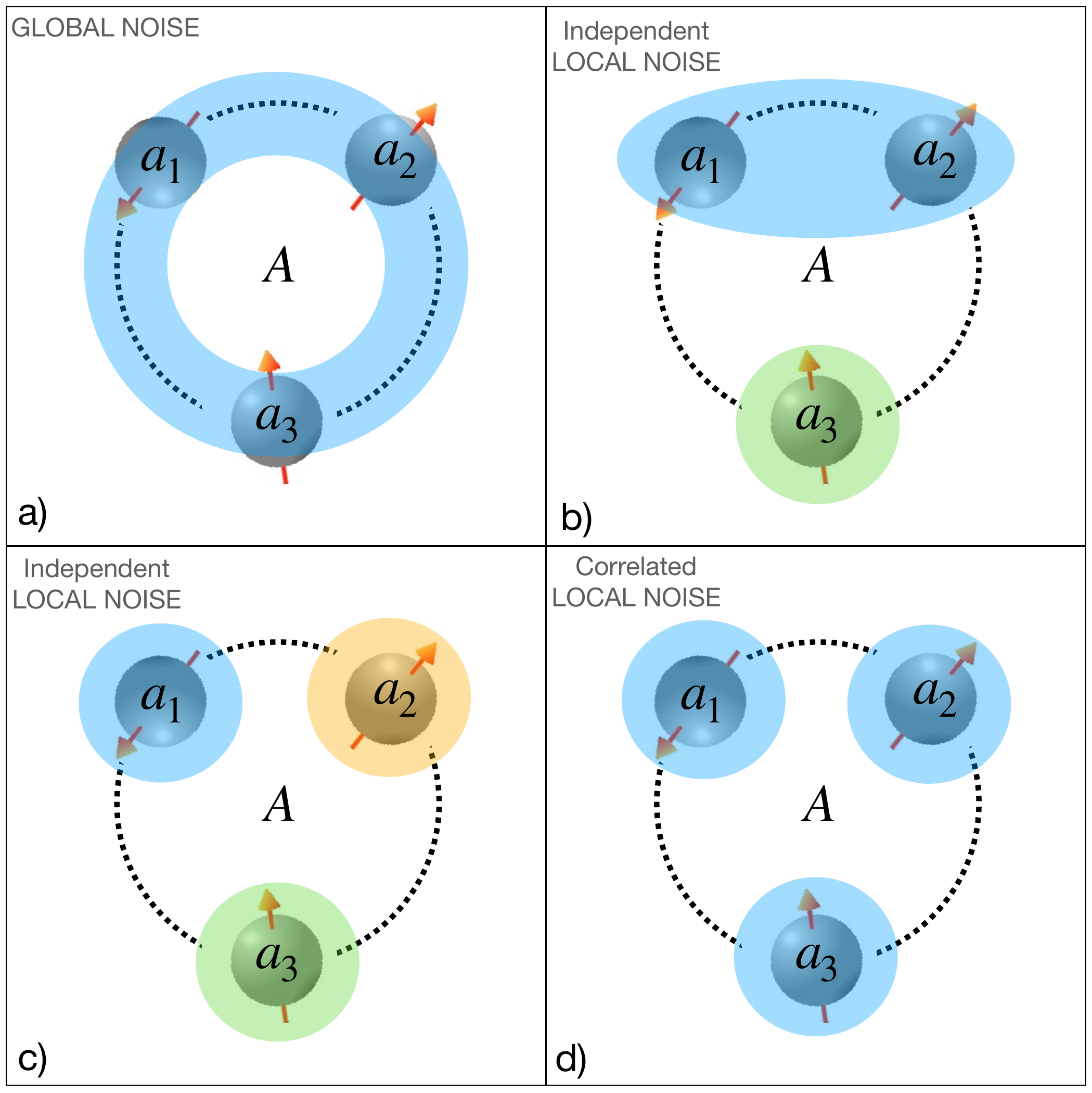}
        \caption{
       \label{figura1}
        Pictorial representation of the noise models considered in this work.
     System $A$ consists of  $n=3$ subsystems ($a_1, a_2, a_3$) subjected to Markovian noise governed by the Lindbladian generator ${\cal L}$~\cite{GKS_Lindblad, L_Lindblad}.
    Panel a) -- Global noise: ${\cal L}$ acts on the entire system (cyan area), potentially introducing spurious correlations. 
    Panel b) -- Independent partially local noise: ${\cal L}$ decomposes into two terms - one acting globally on  $a_1$ and $a_2$ (cyan area), and another on $a_3$ alone (green area).
    Panel c) -- Independent fully local noise scenario: ${\cal L}$ is the sum of three independent terms, each acting separately on
    $a_1$ (cyan area), $a_2$ (red), and $a_3$ (green).
    Panel d) -- Correlated local noise scenario: Similar to c), but with each subsystem affected by the same local Lindbladian term.
        }
\end{figure}

 Noise in quantum systems arises from external disturbances that degrade essential properties like coherence and entanglement~\cite{Breuer2002,Weiss2000,Rivas2011}, both crucial for reliable quantum technologies. A common approach to modeling noise is the Markovian approximation, which assumes memoryless dynamics -- meaning the system's future evolution depends only on its current state, not its past~\cite{GKS_Lindblad, L_Lindblad}. While this simplifies analysis, it remains computationally challenging, especially for large systems with multiple
 components such quantum memories setups where collections of qubits are used to store quantum information over time~\cite{Lvovsky2009,Heshami2016}. 
 Understanding how noise affects these systems is crucial for preserving information integrity. In this paper, we compare different noise models based on how they act on individual subsystems, distinguishing between global and local scenarios~\cite{Gonzalez2017,Cattaneo2019,PhysRevA.102.052208,Zou2024}. In the global case, the dynamical generator ${\cal L}$, which defines the Markov process, operates collectively on the entire system. In contrast, in local scenarios,  ${\cal L}$ is expressed as a sum of local (or partially local) contributions -- see Fig.\ref{figura1}. Rather than focusing on specific cases as usually done in this type of studies, we adopt a statistical approach inspired by 
 Refs.~\cite{PhysRevLett.123.140403,CGDP1}, evaluating different ensembles of Markovian noise models with predefined correlation structures  (global, local, correlated local). 
To measure their impact, we use the Positive Partial Transpose Time (PPTT in brief) -- a metric that quantifies how long entanglement survives as a useful resource under noise~\cite{CGDP1,PhysRevA.99.032307}.
Given a master equation with a dynamical generator ${\cal L}$, the PPTT, denoted as $\tau_{ppt}(\mathcal{L})$, is the minimum time after which the system’s evolution becomes a positive partial transpose (PPT) channel. It can be computed by integrating numerically the master equation. For each ensemble of Lindbladians, we can then derive a probability distribution of PPTT values, providing insight into noise resilience across different scenarios.
According to our findings, when the sampling of different noise scenarios is carried out under controlled and homogeneous conditions that avoid introducing spurious biases, the scenario with the least detrimental impact on the system corresponds to models in which the Lindbladian generator includes global terms
 (Panel a) of Fig.~\ref{figura1}).
In contrast, independent local noise models, where each subsystem is affected by separated (possibly correlated) generators (Panels b), c) and d)),
exhibit a distribution of PPTTs with lower characteristic times, indicating that entanglement tends to decay more rapidly in these cases.
We conclude by emphasizing that a key element in our analysis was the use of the approximate integration method proposed by Cao and Lu~\cite{Cao_Lu}. This approach provides a faster and more efficient alternative to the techniques employed in~\cite{CGDP1} for examining the PPT properties of small quantum systems under random Markovian noise. Notably, it allowed us to reconstruct the PPTT distributions for models with a total dimension $D$ up to~$8$.

The work is organized as follows. In Sec.~\ref{Sect:Markovian_evolution_qsystem}, we introduce the notation and the concept of PPTT for Markovian dynamics, along with PPTT distributions.
In Sec.~\ref{cao-lu}, we review the approximate method proposed by Cao and Lu~\cite{Cao_Lu} for the numerical integration of Markovian processes.
Section~\ref{Sect:comparison} evaluates the performance of the Cao-Lu method in numerically reconstructing the empirical PPTT distribution by sampling the dynamical generators of Markovian evolution. In Sec.~\ref{Sect:PPTT_RandomNoise} we apply this numerical technique to 
reconstruct the  PPTT distributions for medium-sized quantum systems, with dimensions up to $D=8$, that evolves under the action of global noise
and derive an ansatz for higher dimensional systems. 
Building up from the previous results, in Section~\ref{sec:LNvsGN} we finally investigate how the different types of noise correlations of Fig.~\ref{figura1} impact the PPTT distributions of composite quantum systems.
Conclusions and final remarks are provided in Sec.~\ref{Sect:Conclusions}, followed by several technical appendices.

\section{Entanglement degradation in  Markovian quantum noise evolutions}\label{Sect:Markovian_evolution_qsystem}
    Consider a quantum system $A$ of finite dimension~$D$, evolving under of the influence of a Markovian noise
 described by a Gorini-Kossakowski-Sudarshan-Lindblad (GKSL) master equation~\cite{GKS_Lindblad, L_Lindblad}.  
Given  the density matrix $\hat{{\rho}}_A(t)$ of $A$ at time $t$,   its dynamics are governed by
    \begin{equation}\label{masterEQUATION} 
        \dot{\hat{{\rho}}}_A(t) = \gamma {\cal L}
        (\hat{{\rho}}_A(t))\;,
    \end{equation}
    where ${\cal L}$ is the {Lindbladian}  superoperator generating the time evolution. It is defined as
    \begin{equation}
    \mathcal{L}\label{defLL} 
    (\, \cdots \,):= -i k [\hat{H},\, \cdots \,]+  \mathcal{D}_{K}(\, \cdots \,)\;,
     \end{equation} 
 with $\hat{H}$ denoting 
the Hamiltonian component,  and $\mathcal{D}_{K}$   the dissipator,   which accounts for the non-unitary effects induced by the environment.
The dissipative part $\mathcal{D}_K$ can be conveniently expressed as  a sum of independent terms:
\begin{eqnarray}{\label{eq:L_H_and_L_D}}
            &&\mathcal{D}_{K}(\, \cdots \,) \coloneqq \sum_{\ell=1}^{D^2 -1}  \Big(  \hat{L}^{(\ell)}_K(\, \cdots \,) \hat{L}^{(\ell)\dagger}_K \\
            &&\quad  - \frac{1}{2} \left( \hat{L}^{(\ell)\dagger}_K\hat{L}^{(\ell)}_K (\, \cdots \,) + 
           (\, \cdots \,)  \hat{L}^{(\ell)\dagger}_K\hat{L}^{(\ell)}_K \right) \Big)\;,\nonumber 
    \end{eqnarray}
    where the $\hat{L}^{(\ell)}_K$'s are  traceless Lindblad operators associated with distinct dissipative channels.
   These operators satisfy the
    orthogonality relation \begin{eqnarray} \text{Tr}[\hat{L}_{K}^{(\ell)\dagger}\hat{L}_{K}^{(\ell')}]= \lambda_K^{(\ell)} \; \delta_{\ell,\ell'}\;, \end{eqnarray} 
  where $\lambda_K^{(\ell)}$ characterizes the decay rate associated with $\ell$-th channel. 
    In writing Eqs.~(\ref{masterEQUATION}) and (\ref{defLL}), we have adopted natural units by setting  $\hbar = 1$.
    The positive parameters $\gamma$ and $k$ are introduced as tunable coefficients: $\gamma$ 
  set  the overall time-scale of the dynamics, while $k$  controls the  relative strength between the unitary and non unitary contributions
  ~\cite{CGDP1} (in particular  for $k=0$,
   the evolution described by Eq.~(\ref{defLL}) is purely dissipative).
        The subscript $K$ in $\mathcal{D}_{K}$  comes from the fact that, 
        assigning a set $\{\hat{F}_{m}\}_{m=1,\cdots,D^2 -1}$ of traceless and orthonormal matrices with respect to the Hilbert-Schmidt scalar product, i.e.~$\text{Tr}[\hat{F}_{m}^{\dagger}\hat{F}_{m'}]=\delta_{m,m'}$, 
         this term can be also expressed in terms of a $(D^2-1)\times (D^2-1)$ positive semidefinite matrix $K$, called Kossakowski matrix, as:
    \begin{eqnarray}\nonumber
        &&{\cal D}_{K}(\,\cdots\,) = \sum_{m,m'=1}^{D^2 -1}  K_{m,m'} \Big( \hat{F}_{m'}(\,\cdots\,)\hat{F}_{m}^{\dagger} \\
        &&\; - \frac{1}{2}\left( \hat{F}_{m}^{\dagger}\hat{F}_{m'}(\,\cdots\,) +(\,\cdots\,) 
        \hat{F}_{m}^{\dagger}\hat{F}_{m'} \right) \Big),  \label{eq:Dissipator_Kmn}
    \end{eqnarray}
    where $K_{m,m'}$ are elements of $K$. Note that $K\ge0$ is essential to guarantee that the dynamics is trace-preserving and completely positive. Moreover, by diagonalizing $K$ it is possible to obtain the expression of the dissipator appearing in Eq.~\eqref{eq:L_H_and_L_D}. Specifically, one can reconstruct the Lindblad operators as
    \begin{equation}{\label{eq:Lind_op}}
        \hat{L}^{(\ell)}_K := \sqrt{\lambda^{{(\ell)}}_K} \sum_{m}   U^{(m,\ell)}_K \hat{F}_{m},
    \end{equation}
    expressing them in terms of the eigenvalues $\lambda^{{(\ell)}}_K$ of $K$ and of matrix elements  $U^{(m,\ell)}_K$ which define  its eigenvectors through the spectral decomposition $K_{m,m'} = \sum_{\ell} U^{(m,\ell)}_K U^{(m',\ell)*}_K  \lambda^{{(\ell)}}_K$.
We emphasize   that the rank  of the matrix $K$ determines
the number of independent noise channels (i.e., the total number of distinct noise generators acting on system $A$),
 while 
its trace 
sets the arithmetic mean value of the dissipative rate constants $\lambda_K^{(\ell)}$, 
\begin{eqnarray}\label{HILBERTS}
\langle \lambda \rangle :=\frac{\sum_{\ell=1}^{D^2-1}\lambda^{{(\ell)}}_K }{D^2-1}  = \frac{\mbox{Tr} [ K ] }{D^2-1} \;,
\end{eqnarray} 
 (the identity stems from
~(\ref{eq:Lind_op}) and from the orthogonality conditions of the operators $\hat{F}_{m}$).

\subsection{Positive partial transpose time of GKSL generators} \label{sec:pptt} 
    The formal integration  of Eq.~{\eqref{masterEQUATION}} on the time interval $[t_0, t]$ is provided by the  Linear, Completely Positive, Trace Preserving 
  (LCPTP) dynamical map \begin{equation}{\label{eq:solution_Lindblad_eq}}
        {\Phi}_{(t_0,t)}:=e^{\gamma(t -t_0)\mathcal{L}}\;,
    \end{equation}  
    which inherits from the GKSL generator  ${\mathcal{L}}$ 
an implicit dependence upon the parameter $k$ and the Kossakowski matrix $K$.
     Specifically given 
     $\hat{\rho}_{A}(t_0)$ the input state of $A$ 
     at time $t=t_0$, 
     its evolved counterpart at time $t\geq t_0$ 
     under the action of the generator ${\mathcal{L}}$
     can be expressed as  
       $\hat{\rho}_{A}(t) = \Phi_{(t_0,t)}(\hat{\rho}_{A}(t_0))$.
   By construction the maps $\Phi_{(t_0,t)}$ are time-homogeneous, meaning that they
    only depend upon the length of the
    time interval $[t_0,t]$, and form  a dynamical semigroup, i.e. 
    \begin{eqnarray} \left\{ \begin{array}{l}
      \Phi_{(t_0,t)}=\Phi_{t-t_0} \;, \\ \\
      \Phi_{t}\circ\Phi_{t'}=  \Phi_{t+t'} \;, \qquad \Phi_{0} = \mathcal{I}\;, \end{array} 
      \right.  \label{dinsemi} 
    \end{eqnarray} 
    with $\mathcal{I}$ being the identity channel.
    We also recall that, given $B$ an auxiliary system which has the same dimension of $A$, 
   the Choi-Jamiołkowski (CJ) state~\cite{Jam,Choi} of the transformation $\Phi_{(t_0,t)}$ 
   is defined as the evolution of a maximally entangled state 
    $\ket{\Psi_{\max}}_{AB}$ of $A$ and $B$, 
     under the action of channel, i.e.  
    \begin{equation}\label{defCJ} 
    \hat{\rho}_{AB}^{\Phi_{(t_0,t)}} \coloneqq (\Phi_{(t_0,t)}\otimes \mathcal{I}_{B}) (\ket{\Psi_{\max}}_{AB} \bra{\Psi_{\max}}),
\end{equation}
    with $\mathcal{I}_{B}$ indicating the identity map on $B$. 
  The state $\hat{\rho}_{AB}^{\Phi_{(t_0,t)}}$ provides an exhaustive characterization of $\Phi_{(t_0,t)}$. In particular, as 
    discussed in Refs.~\cite{CGDP1,PhysRevA.99.032307}  a {\it bona fide} indicator of the noise level induced by  the dynamical semigroup $\Phi_{t}$ on the system $A$ is provided by the PPTT $\tau_{ppt}(\mathcal{L})$ of the GKSL generator ${\cal L}$. This corresponds 
    to the first  time after which the associated dynamical map
 $\Phi_{t}$  becomes 
 a positive partial transpose (PPT) channel and
 bounds the maximum time we can wait before 
 $A$ becomes useless as a resource for the 
 generation of maximally entangled states via local operations and classical communication (LOCC), even when multiple copies of the same input configuration are available.
 Most importantly, at variance with other noise indicators (like for instance the minimum time at which $A$ losses all its entanglement), once the  dynamics of the system is solved, the value of  $\tau_{ppt}(\mathcal{L})$
 can be computed 
    as the minimum positive value of  $t$, for which the entanglement negativity~\cite{Vidal}  
    of 
the  CJ state~(\ref{defCJ}) reaches zero, i.e. 
\begin{equation}\label{eq:est_time}
    \tau_{ppt}(\mathcal{L}
    ) \coloneqq \min\{ t\ge0 \text{ s.t. } 
    \mathcal{N}(\hat{\rho}_{AB}^{\Phi_{t}})= 0\},
\end{equation}
where indicating with  $\top_B$  the partial transpose w.r.t. to an orthonormal basis of $B$ and with
$\| \cdots \|_1$ for the trace norm one has $\mathcal{N}(\hat{\rho}_{AB}^{\Phi_{t}}):=
( \| (\hat{\rho}_{AB}^{\Phi_{t}})^{\top_B}\|_1-1)/2$.

As mentioned in the introduction of the manuscript, instead of focusing on specific instances of the noise model, 
 we aim to provide a statistical characterization of PPTTs for ensembles of the dynamical generator ${\mathcal{L}}$. 
Specifically, given $d \mu_D({\mathcal{L}})$ a probability measure on  the set $\mathfrak{L}$ of the GKSL generators of $A$ that select a specific structure of the noise,  we study the induced probability density function $P_{ppt}(\tau)$ to find a Lindbladian  ${\cal L}$ with $\tau_{ppt}({\mathcal{L}})=\tau$. For any assigned
infinitesimal interval $[ \tau,\tau+\delta \tau[$,  $P_{ppt}(\tau)$ is 
 formally identified by the expression \begin{eqnarray} 
P_{ppt}(\tau) d\tau &: =& \mbox{Prob}\Big( \tau_{ppt}({\mathcal{L}}) \in[ \tau,\tau+\delta \tau[ \Big) \nonumber \\
&=& \int_{\mathfrak{L}[{\tau}, \delta \tau]} d\mu_D({\mathcal{L}})  \;, \label{mpodef} 
\end{eqnarray}
with $\mathfrak{L}[{\tau}, \delta \tau]$ the subset of $\mathfrak{L}$ of $A$ which contains all the Lindbladian with PPTT values in 
 $[ \tau,\tau+\delta \tau[$.
 Furthermore we also consider
the cumulative distribution function associated to $P_{ppt}(\tau)$, i.e. 
\begin{eqnarray} \label{cumu} 
\bar{P}_{ppt}(T) &: =& \mbox{Prob}\Big( \tau_{ppt}({\mathcal{L}}) \leq T\Big) \nonumber \\
&=&\int_0^T d\tau P_{ppt}(\tau)\;,
\end{eqnarray}
which, 
as discussed in~\cite{CGDP1}, represents the probability that a maximally entangled state of the joint system $AB$ transitions to a PPT state by time $T$, accounting for the uncertainty in the dynamics of subsystem $A$ induced by the selected 
measure $d\mu_D({\cal L})$. 
In this manuscript, we focus on purely dissipative models  (i.e. with $k=0$) 
and employ techniques from random matrix theory to sample the corresponding dissipator 
 $\mathcal{D}_K$ via a measure $d\mu^{(\xi_D,r_D)}_D(K)$ defined for the Kossakowski matrix $K$ [The only exception to this setting is presented in Appendix~\ref{sec:comparison}, where we include results for the case $k=1$].
Specifically, we use $d\mu^{(\xi_D,r_D)}_D(K)$ to sample  $K$ 
from the  ensemble of $(D^2-1)\times (D^2-1)$ Wishart matrices, constraining its trace to a positive scaling function
 $\xi_D$
of the system dimension $D$
  \begin{eqnarray} \label{constraints} 
   \mbox{Tr}[K]=\xi_D\;, 
  \end{eqnarray} 
  and imposing an upper bound $r_D \leq D^2-1$ on its  rank,
    \begin{eqnarray} \label{constraints2} 
   \mbox{rank}[K] \leq r_D\;. 
  \end{eqnarray} 
Concretely, $K$ is generated as 
  $\xi_D \frac{G^\dag G}{\mbox{Tr}[ G^\dag G]}$
where $G$ is  a complex $r_D\times (D^2-1)$ Ginibre matrix with independent complex Gaussian entries~\cite{PhysRevLett.123.140403,Zyczkowski2011,Pastur2011}.
Adopting these techniques previous investigations~\cite{CGDP1} have explored the $P_{ppt}(\tau)$ and $\bar{P}_{ppt}(T)$ 
 in bipartite systems of qubits $(D=2)$ and qutrits $(D=3)$.
  Extending this analysis to systems of higher dimensionality may reveal whether the trends observed in simpler cases generalize to more complex systems. However, such an extension poses significant computational challenges due to the increasing time required to compute the  CJ density matrix $\hat{\rho}_{AB}^{\Phi_{(t_0,t)}}$, as dictated by the solution to the Lindblad master equation.

    \section{Cao-Lu master equation integration method} \label{cao-lu} 
    
  A direct method to determine  the CJ state~(\ref{defCJ}) and the corresponding PPTT of a GKSL generator, is to 
 work  in Liouville representation and
  compute the exponential of the matrix which describes ${\mathcal{L}}$ via diagonalization of the latter. 
This is what we will refer to as the \textit{standard method} to solve the master equation~(\ref{masterEQUATION}). Such an  approach is certainly valid when one wants to study small systems, as it provides great accuracy without consuming too much time resources in a numerical computation. However, when one is interested in studying larger systems, it may be expensive to directly compute matrix exponential of super-operators. 
  Therefore, it might be useful to have approximated methods to simulate this dynamics, which also preserve the properties that a density matrices must have, namely it must be positive semi-definite and with unit trace.
    A potentially promising scheme has been proposed by Cao and Lu~\cite{Cao_Lu}.
    This approach, which we shall refer to as the \textit{Cao-Lu method}, bypasses the need for diagonalization of the GKSL generator, relying instead on matrix multiplications, which may result in a reduction in computational time.
The starting point of this analysis is the decomposition of 
Strinbach, Garraway, and Knight~\cite{Strinbach} where $\Phi_{(t_0,t)}$ 
 is expressed as an infinite sum of terms 
each representing a 
 trajectory in which  the system $A$ experiences 
a discrete number of sharp transitions ({\it quantum jumps}) induced by the super-operator
  \begin{equation}
    {\label{eq:CP_splitting}}
         \mathcal{Q}(\, \cdots \,) \coloneqq  \sum_{\ell=1}^{D^2 -1} \hat{L}^{(\ell)}_K(\, \cdots \,) \hat{L}^{(\ell)\dagger}_K\;.
    \end{equation}
Specifically  introducing  the  (non-Hermitian)
{effective Hamiltonian} 
    \begin{equation} \label{Heff} 
        \hat{H}_{\text{eff}} \coloneqq k \hat{H} + \frac{1}{2i}\sum_{\ell=1}^{D^2 -1}\hat{L}^{(\ell)\dagger}_{K} \hat{L}^{(\ell)}_{K}\;, 
    \end{equation}
    and the super-operator 
        \begin{equation}
    {\label{eq:CP_splitting1}}
     \mathcal{J}(\, \cdots \,) \coloneqq \hat{J}(\, \cdots \,) + (\, \cdots \,) \hat{J}^{\dagger},
     \qquad \hat{J} \coloneqq -i \hat{H}_{\text{eff}}\;,
    \end{equation}
one can rewrite the r.h.s. of Eq.~(\ref{eq:solution_Lindblad_eq}) as
         \begin{equation}{\label{eq:rho_integral_form}}
          \Phi_{(t_0,t)}  = \sum_{m=0}^\infty  \mathcal{N}^{(m)}_{(t_0,t)} 
  \;, 
    \end{equation}
with the  $\mathcal{N}^{(m)}_{(t_0,t)}$'s  recursively defined by the 
 identities
         \begin{eqnarray}{\label{eq:rho_integral_form1}}
          \left\{ \begin{array}{ll}  \mathcal{N}^{(0)}_{(t_0,t)}  &= e^{\gamma(t-t_0) \mathcal{J}}\;, \\ \\
            \mathcal{N}^{(m+1)}_{(t_0,t)}  
            &= \gamma \int_{t_0}^{t} dt' 
            e^{\gamma(t-t') \mathcal{J}}\circ \mathcal{Q} \circ \mathcal{N}^{(m)}_{(t_0,t')} \;.\end{array}
           \right.
    \end{eqnarray}
Notice that in the limit of  short time intervals (i.e. $\gamma (t-t_0)\ll 1$) the leading contribution of the  super-operator $\mathcal{N}^{(m)}_{(t_0,t)}$ scales  as $(\gamma (t-t_0))^m$. Furthermore 
 as in the case of $\Phi_{(t_0,t)}$, the $\mathcal{N}^{(m)}_{(t_0,t)}$'s are time-homogenous, 
    \begin{eqnarray}\mathcal{N}^{(m)}_{(t_0,t)} = \mathcal{N}^{(m)}_{t-t_0}\;,
    \end{eqnarray} 
   even though they typically lack of the semigroup property due to the non-commutativity of $\mathcal{Q}$ and $\mathcal{J}$. 
 Most importantly one can also observe that, while not necessarily trace preserving, all the  $\mathcal{N}^{(m)}_{(t_0,t)}$ are 
explicitly Completely Positive (CP)~\cite{Cao_Lu}. 
This implies that truncations of the series (\ref{eq:rho_integral_form}) provide well-behaved approximations of the dynamical map 
 $\Phi_{(t_0,t)}$. Specifically, incorporating all contributions involving at most $M$ quantum jumps within 
a given time interval $[t,t+\Delta t]$  of length $\Delta t < 1/\gamma$, 
one can write   \begin{equation}{\label{eq:rho_integral_form1M}}
     \Phi_{(t,t+\Delta t)}  =  \Phi_{\Delta t}  =   \sum_{m=0}^{M}  \mathcal{N}^{(m)}_{\Delta t}  +  \mathcal{O}((\gamma{\Delta t})^{M+1})
  \;.
    \end{equation}
Notice that  evaluating the r.h.s. of (\ref{eq:rho_integral_form1M}) 
 still requires computing the matrix exponentials $e^{\gamma t \mathcal{J}}$ appearing in  Eq.~\eqref{eq:rho_integral_form1}, which, similarly to what happens for the standard
 integration method, becomes increasingly challenging as the dimension of $A$ grows.
  This is where Cao and Lu make a crucial observation: by leveraging the fact that $e^{\gamma t \mathcal{J}}$ admits a Kraus representation in terms of the operators $e^{\gamma t \hat{J}}$, they demonstrated that the following identity holds~\cite{Cao_Lu}:
    \begin{equation}{\label{eq:exp_approx}}
        e^{\gamma t \mathcal{J}} = \mathcal{J}^{(\ell)}_{ t } + \mathcal{O}((\gamma t)^{\ell+1})\;, 
    \end{equation}
    where for $\ell$ integer one defines 
    \begin{equation}\label{defJ} 
        \mathcal{J}^{(\ell)}_{ t }\coloneqq
      \mathcal{K}_{\hat{A}}\;, \qquad \hat{A}:= 
 \sum_{\alpha=0}^{\ell}\tfrac{\left(
 \gamma t  \hat{J}\right)^{\alpha}}{\alpha!}\;, 
       \end{equation}
    with   $\mathcal{K}_{\hat{A}}$ the super-operator 
    \begin{eqnarray} \label{DefKA} \mathcal{K}_{\hat{A}}(\, \cdots \,)\coloneqq \hat{A} (\, \cdots \,) \hat{A}^{\dagger}\;.\end{eqnarray} 
   Substituting 
\eqref{eq:exp_approx} in Eq.~\eqref{eq:rho_integral_form1} we can hence conclude that given $\ell$ and $m$ integers such that $\ell \geq m$, one has 
  \begin{equation}{\label{eq:rho_integral_formapp}}
            \mathcal{N}^{(m)}_{\Delta t}  =    \overline{\mathcal{N}}^{(m;\ell)}_{\Delta t} + \mathcal{O}((\gamma \Delta t)^{\ell+1} )\;, 
      \end{equation}
   with $\overline{\mathcal{N}}^{(m;\ell)}_{\Delta t}$ defined recursively via the identities 
    \begin{equation}{\label{eq:rho_integral_form2}}
         \!\! \left\{ \begin{array}{ll}  \overline{\mathcal{N}}^{(0;\ell)}_{\Delta t}  &:=  \mathcal{J}_{\Delta t}^{(\ell)}\;, \qquad \forall \ell\geq 0\\ \\
            \overline{\mathcal{N}}^{(m+1;\ell)}_{\Delta t}  
            &:=\gamma \int_{0}^{\Delta t} dt' 
            \mathcal{J}_{\Delta t-t'}^{(\ell-m-1)} \circ \mathcal{Q} \circ  \overline{\mathcal{N}}^{(m;\ell-1)}_{t'}.
           \end{array}
           \right.
    \end{equation}
    Therefore setting $\ell = M$, from Eq.~(\ref{eq:rho_integral_form1M}) we finally get 
    \begin{equation}{\label{eq:rho_integral_form1Mnew}}
         \Phi_{\Delta t}  = \sum_{m=0}^{M}  \overline{\mathcal{N}}^{(m; M)}_{\Delta t}  +  \mathcal{O}((M+1)({\gamma \Delta t})^{M+1})\;,
    \end{equation}
where now all the terms in the r.h.s. are in the Kraus form, thanks to Eqs.\eqref{eq:exp_approx}--(\ref{DefKA}). The conceptual procedure to obtain a proper approximation scheme for $\Phi_{\Delta t}$ is complete. In order to compute numerically the integrals that appear in Eq.~(\ref{eq:rho_integral_form2}), one can use appropriate quadrature methods, with the only constraint that they do not modify the order of approximation of the scheme.
In this paper we will focus on the midpoint rule, given its simplicity, the reduced computational cost, and the fact that it evaluates the integrand at the most representative point of each interval.
Accordingly if one considers as maximum number of quantum jumps allowed $M=2$, Eq.~(\ref{eq:rho_integral_form1Mnew})  can be expressed as
\begin{equation}\label{primaapprox}
\Phi_{\Delta t}= \mathcal{A}_{\Delta t}^{(\text{2,MP})}+ \mathcal{O}((
\gamma \Delta t)^{3} )\;,
\end{equation}
with 
   \begin{eqnarray}\label{eq:midpoint_rule_Kraus_scheme}
          \mathcal{A}_{\Delta t}^{(\text{2,MP})} &\coloneqq&  \mathcal{K}_{\hat{A}''}  \\ &&\!\!+(\gamma\Delta t) \; \mathcal{K}_{\hat{A}'}\circ \mathcal{Q}\circ \mathcal{K}_{\hat{A}'} \nonumber 
          + \frac{(\gamma \Delta t)^2}{2}\mathcal{Q}^2\;,
    \end{eqnarray}
where $\mathcal{K}_{\hat{A}'}$ and $\mathcal{K}_{\hat{A}''}$ are defined as in Eq.~(\ref{DefKA}) with the operator $\hat{A}$ replaced by 
\begin{eqnarray}
\hat{A}'&:=&\hat{\mathbb{1}} + \frac{(-i\hat{H}_{\text{eff}}) (\gamma \Delta t)}{2}\;,\\
\hat{A}''&:=& \hat{\mathbb{1}}+ (-i \hat{H}_{\text{eff}}) (\gamma \Delta t) + \frac{(-i \hat{H}_{\text{eff}})^2  (\gamma \Delta t)^2}{2}\;. \nonumber
\end{eqnarray} 
To extend Eq.~(\ref{primaapprox}) to  time intervals  $[t_0,t]$ of  length $T:=t-t_0$ which is not shorter than $1/\gamma$,  we can  invoke 
the dynamical semigroup property~(\ref{dinsemi}) splitting $\Phi_{(t_0,t)}$ into $N$ small enough intervals
of length $\Delta t: =T/N < 1/ \gamma$ and use (\ref{eq:midpoint_rule_Kraus_scheme}) on each of them,
\begin{eqnarray} \label{decCAO-LU}
\Phi_{(t_0,t)}&= & \left(\Phi_{\Delta t}\right)^{N} =  \left( \mathcal{A}_{\Delta t}^{(\text{2,MP})}+ \mathcal{O}((
\gamma \Delta t)^{3} )\right)^N \nonumber \\
&=& \left(\mathcal{A}_{\Delta t}^{(\text{2,MP})}\right)^N + \mathcal{O}(N(
\gamma \Delta t)^{3} )\;.
\end{eqnarray}

\section{Performance Evaluation}\label{Sect:comparison}

In a previous work~\cite{CGDP1}, some of us 
obtained the PPTT distributions for bipartite qubit and qutrit systems by directly integrating the associated Lindblad master equations. However, as noted earlier,  this approach is  computationally expensive.
In particular, computing the
 CJ state of the system at a given time $t$ requires evaluating the exponential super-operator $e^{\gamma t{\mathcal{L}}}$.
 This process becomes increasingly demanding in terms of both time and computational
 resources as the system dimension grows, making direct numerical simulation impractical for larger systems. 
 A key computational bottleneck in this standard approach is the need to diagonalize the matrix associated with the generator  $\mathcal{L}$, which has a dimension of $D^2$, scaling quadratically with the dimension $D$ of the subsystem $A$. 
Given that diagonalization -- typically performed using the QR algorithm -- requires a number of iterations on the order of $\mathcal{O}(M^3)$, where $M$ is the matrix size, the overall computational complexity of this method is expected to be at least $\mathcal{O}(D^6)$.
The goal of this section is to assess whether the alternative Cao-Lu method, outlined in the previous section, can reduce the computational costs associated with computing the PPTT distribution~(\ref{mpodef}) for medium-sized systems -- i.e., those larger than the ones analyzed in Ref.~\cite{CGDP1}.
We start observing  to evaluate the action of the super-operator $\mathcal{A}_{\Delta t}^{(\text{2,MP})}$ on the density matrix of the composite system $A$ and $B$, 
one must compute Kronecker products and standard matrix multiplications. 
Calculating the Kronecker product between two matrices of dimensions $n$ and $m$ has a computational complexity of ${\cal O}(n^2 m^2)$. In our case, this results in a complexity of ${\cal O}(D^4)$. However, the subsequent matrix multiplication has a higher computational complexity of ${\cal O}(D^6)$\footnote{The most commonly used algorithm for matrix multiplication is the Strassen algorithm, which has a complexity of approximately $\mathcal{O}(M^{2.81})$. However, for large matrices, the Strassen algorithm is not usually used due to its high constant factor and it's more common to use the standard matrix multiplication algorithm, which has a complexity of $\mathcal{O}(M^{3})$ for two $M\times M$ matrices. In Python, the complexity of matrix multiplication in libraries like \textit{numpy} is typically $\mathcal{O}(M^{3})$ for large matrices.}, leading to an overall computational cost of the same order as the standard method.
That said, in terms of constant factors, standard matrix multiplication often benefits from highly optimized implementations, whereas the QR algorithm typically has higher overhead, especially for large matrices. This is due to the additional computations required for finding eigenvalues and eigenvectors. 
Thus, the Cao-Lu algorithm is expected to offer a more efficient approach for computing PPTT distributions, particularly for medium-sized bipartite systems. To test this hypothesis, we implemented all relevant processes in Python using the \textit{numpy} library, which takes advantage of highly optimized low-level operations. The results, presented in Appendix~\ref{sec:comparison}, support this hypothesis. However, the exact performance comparison between the two methods also depends on factors such as cache behavior, memory access patterns, and CPU architecture.

Let us now analyze the expected error in estimating the PPTT when using the Cao-Lu method.
From Eq.~\eqref{decCAO-LU}, we know that after $N$ time steps of lenght $\Delta t<1/\gamma$, the evolved CJ state 
$\hat{\rho}_{AB}^{\Phi_{T=N\Delta t}}$ is determined up to an error of the order of $\mathcal{O}(N d x^{3})$ which we expressed in terms of the increment $d x=\gamma \Delta t $ of the dimensionless parameter \begin{eqnarray} x := \gamma t\;,\label{increment}\end{eqnarray}  
which measures the time coordinate in $(1/\gamma)$-units.
We track this evolution over time until the PPT condition is reached at
$T=\tau_{ppt}({\cal L})$, so that 
 the accumulated error on the final state will be of order $\mathcal{O}(f(D) dx^{2})$, with 
$f(D)$ depends on the scaling of 
$\gamma \tau_{ppt}({\cal L})$ w.r.t. $D$. Empirical analysis shows that  $f(D)$ scales as $\log_2(D)$, meaning the final-state error follows  $\mathcal{O}(\log_2(D) d x^{2})$.
Once $\hat{\rho}_{AB}^{\Phi_{T}}$ is computed, the next step is to determine its entanglement negativity, given by 
\begin{eqnarray} \mathcal{N}(\hat{\rho}_{AB}^{\Phi_{T}})=\frac{1}{2}\sum_{l=1}^{D^2}(|\lambda_{l}|-\lambda_{l})\;,\end{eqnarray} 
where $\{\lambda_{l}\}_{l=1,\dots,D^2}$ are the eigenvalues of $(\hat{\rho}_{AB}^{\Phi_{T}})^{\top_B}$ (note that $\mathcal{N} = 0$ if and only if all the eigenvalues $\lambda_l$ are non-negative).
The partial transposition itself is error-free, as it only involves rearranging matrix elements. However, the diagonalization step introduces machine precision errors of approximately $\mathcal{O}(10^{-16})$. 
The dominant source of error in the eigenvalues
 $\{\lambda_{l}\}_{l=1,\dots,D^2}$ thus comes from the inaccuracy of
 the CJ state,  leading to an error scaling as $\mathcal{O}(\log_2(D) dx^{2})$. Since eigenvalues are continuous functions of matrix elements, their errors generally scale in the same order as the matrix perturbation, making this assumption reasonable.
Finally, we consider how this error propagates in the estimation of entanglement negativity. 
Each eigenvalue $\lambda_l$
is known up to an error of $\mathcal{O}(\log_2(D)dx^{2})$. 
When summing over all $D^2$ terms, these individual uncertainties accumulate. Since each term contributes independently, the total error scales as ${\cal O}(D^2 \log_2(D) \, dx^2)$. For larger values of $D$, this error may become significant, potentially leading to a situation where the entanglement negativity of the CJ state is perceived as zero even when it is actually nonzero.

\section{Reconstructing the PPTT distributions }\label{Sect:PPTT_RandomNoise}

In this section we present the empirical PPTT distributions  obtained using the Cao-Lu method for dimensions 
that range from $2$ up to $8$ and use the results to extrapolate a general trend for arbitrary large $D$ values. 
For the sake of simplicity we focus on the case of purely dissipative GKSL generators setting 
 $k=0$    in Eq.~(\ref{defLL}), and 
 express the results 
measuring the time coordinate in 
$(1/\gamma)$-units
 through 
 the identities~\cite{CGDP1}
\begin{eqnarray}\label{IMPO}  
\begin{cases} 
P_{ppt}(\tau)  &= {\gamma} {\mathbf P}_{ppt}(\gamma
\tau) \;, \\\\ 
\bar{P}_{ppt}(T)  &=  \bar{\mathbf P}_{ppt}(\gamma
T) \;, 
\end{cases} 
\end{eqnarray} 
with ${\mathbf P}_{ppt}(x)$ and $\bar{\mathbf P}_{ppt}(x):=  \int_0^x dx'{\mathbf P}_{ppt}(x')$ being the rescaled PPTTs distributions~\cite{CGDP1}.

\begin{figure*}\label{fig:prob_distrib_and_cdf_N_no_fit}
    \subfloat[][]{\includegraphics[scale=0.225]{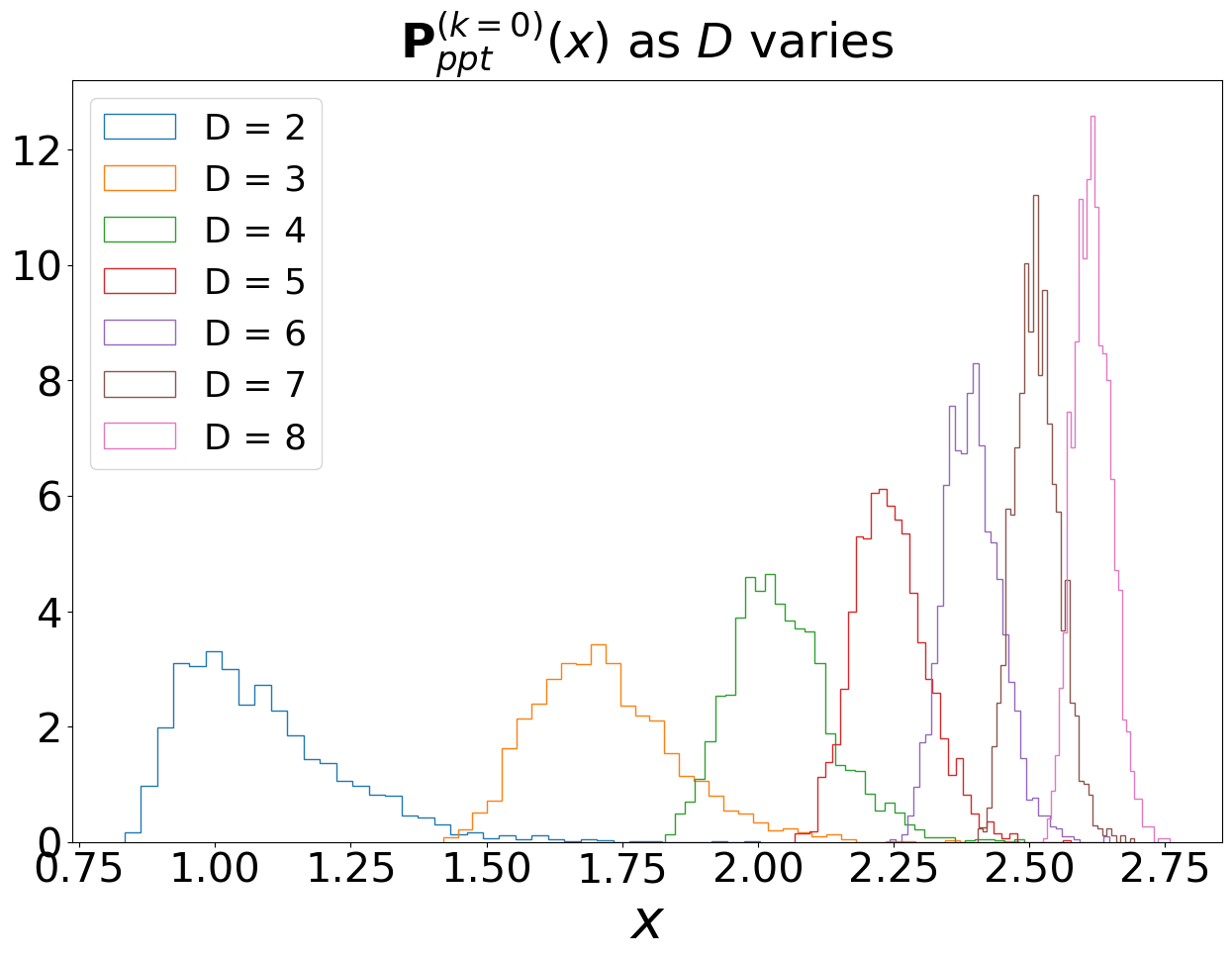}\label{fig:PPTT_distrib}} \quad
    \subfloat[][]{\includegraphics[scale=0.225]{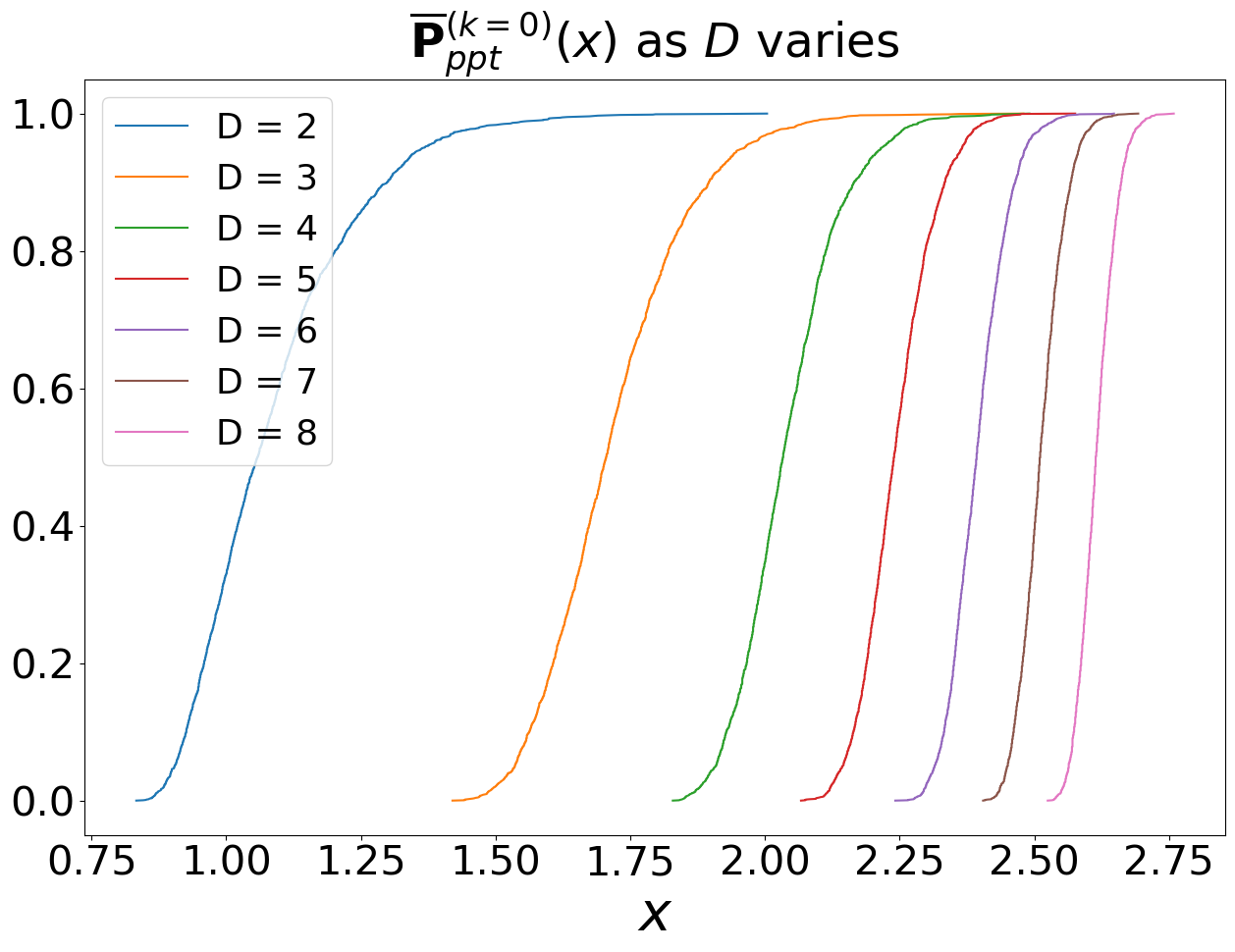}\label{fig:PPTT_cdf}}
    \caption{Empirical rescaled PPTT distributions ${\mathbf P}_{ppt}(x)$ (Panel (\ref{fig:PPTT_distrib})) and their cumulative counterpart $\overline{\mathbf P}_{ppt}(x)$  (Panel (\ref{fig:PPTT_cdf})) for increasing dimension $D$ of one of the two subsystems, obtained considering $2000$ samplings of the rescaled Lindblad generator $\mathcal{L}$ and a time step $dx=0.001$. The plots refer to the case of $k=0$ (purely dissipative models), sampling the $(D^2-1)\times (D^2-1)$ Kossakowski matrix $K$ on the Wishart ensemble with  $\xi_D=D$ and $r_D=D^2-1$ (unrestricted rank value).}
    \label{fig:PPTT_distrib_increasing_N}
\end{figure*}

\subsection{Empirical distributions for not-so-small systems} \label{sec:emp} 
Here, we report the empirical distributions ${\mathbf P}_{ppt}(x)$ and their cumulative counterpart
$\bar{\mathbf P}_{ppt}(x)$ for values of  $D$ up to $8$. These distributions were obtained through numerical sampling of the 
dissipator ${\cal D}_K$ as detailed at the end of Sec.~\ref{sec:pptt} 
under  the normalization condition   \begin{eqnarray}\label{canonical} \xi_D= D\;,  \end{eqnarray} 
as specified in Eq.~(\ref{constraints}), posing no restriction on the rank of the matrix $K$ setting 
\begin{equation} 
r_D = D^2-1\;.\label{rankmax} 
\end{equation} 
 This choice, referred to as the {\it canonical normalization condition}, was first introduced in Ref.~\cite{PhysRevLett.123.140403} and later adopted in~\cite{CGDP1}.
 It offers a convenient framework for presenting the final results, as it ensures a clear separation of the distributions for different values of 
$D$. Finally, as explicitly demonstrated in Sec.\ref{sec:ansatz}, the PPTT distributions corresponding to alternative choices of the function $\xi_D$ can be obtained from the one associated with the canonical normalization condition in Eq.~(\ref{canonical}) via a simple rescaling procedure.
 The calculations were performed over $2000$ iterations, corresponding to the number of random samplings $\mathcal{L}$,
 with a time step increment of $d x=10^{-3}$.
 
 The results are presented in Fig.~\ref{fig:PPTT_distrib_increasing_N}, which illustrate that, for the chosen sampling method, the 
 PPTT distributions exhibit increasing values as $D$ increases. Moreover, the distributions become progressively more concentrated around the mean, indicating that the loss of entanglement (whether partial or complete) occurs within a more predictable time window.
In Appendix~\ref{sec:fits}, we demonstrate that the three-parameter Gamma and Lognormal distributions -- previously shown to provide a good fit for the qubit $(D=2)$ and qutrit $(D=3)$  cases~\cite{CGDP1} -- also effectively approximate the distributions observed in this study.

\begin{figure*}[htbp] % * per farlo estendere su entrambe le colonne
	\centering
	\begin{tabular}{cc}
		% Prima riga
		\subfloat[]{\includegraphics[width=0.45\textwidth]{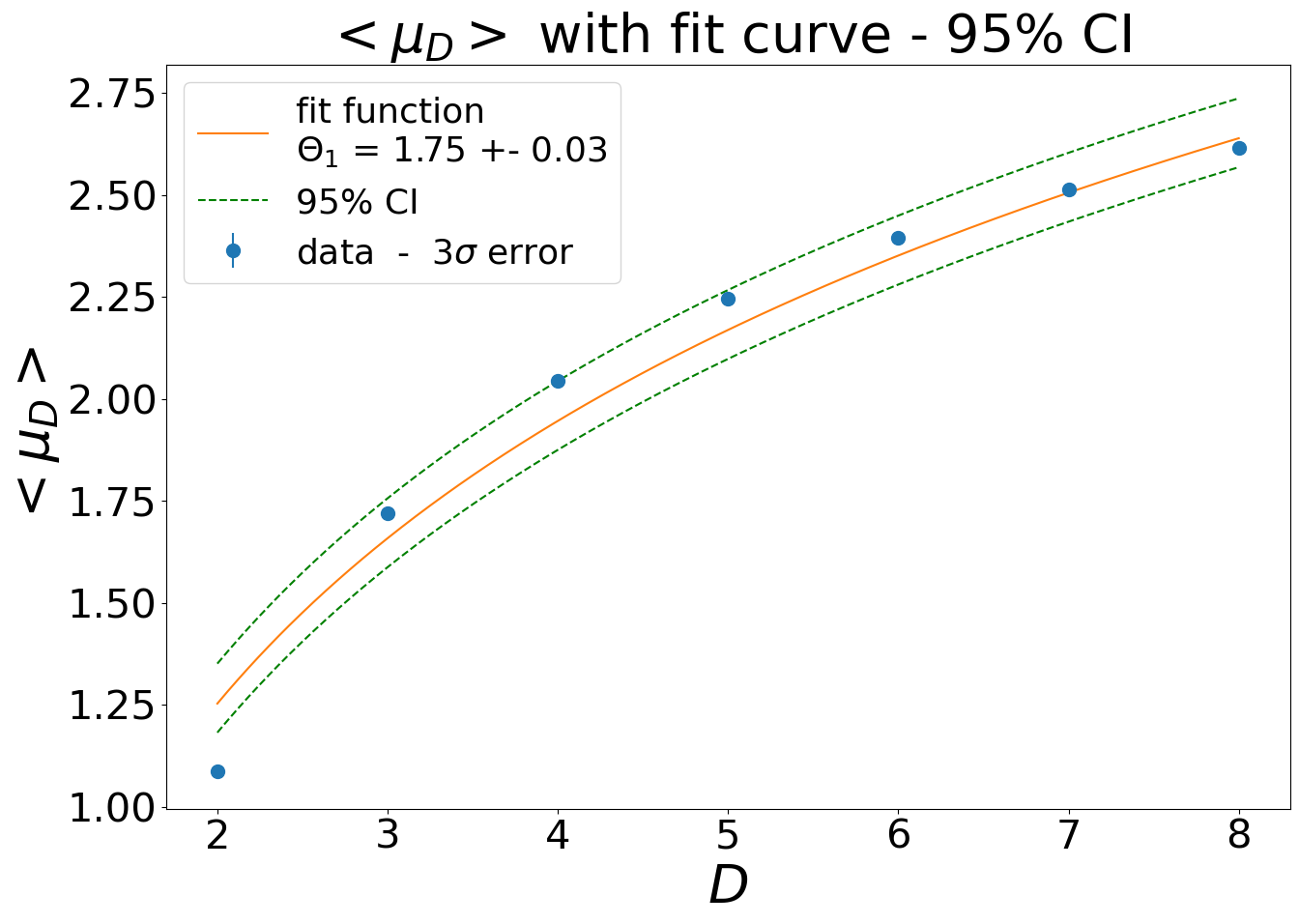}\label{fig:Fit_mean_N}} $\qquad$ & 
		\subfloat[]{\includegraphics[width=0.45\textwidth]{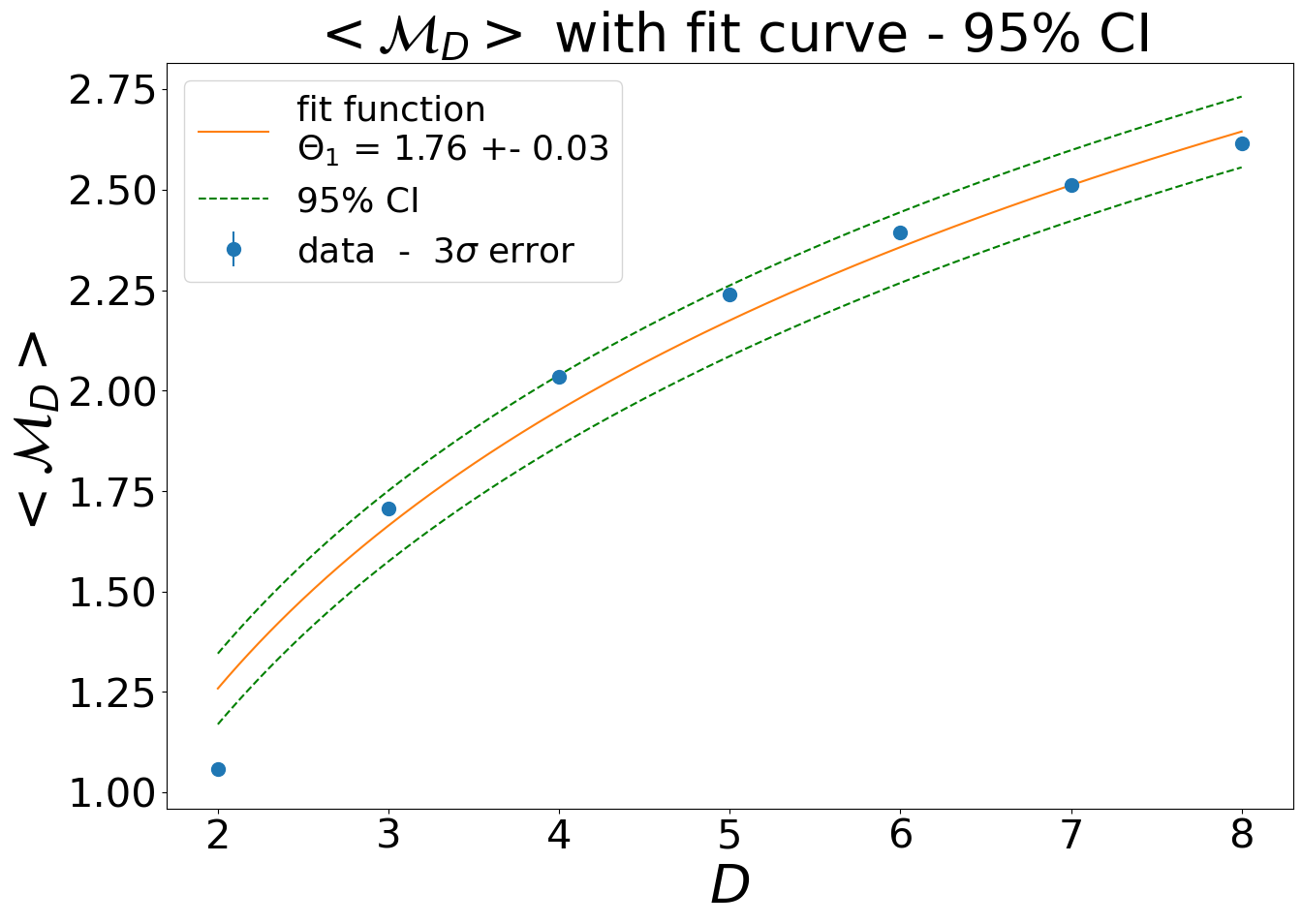}\label{fig:Fit_median_N}} \\
		
		% Seconda riga
		\subfloat[]{\includegraphics[width=0.45\textwidth]{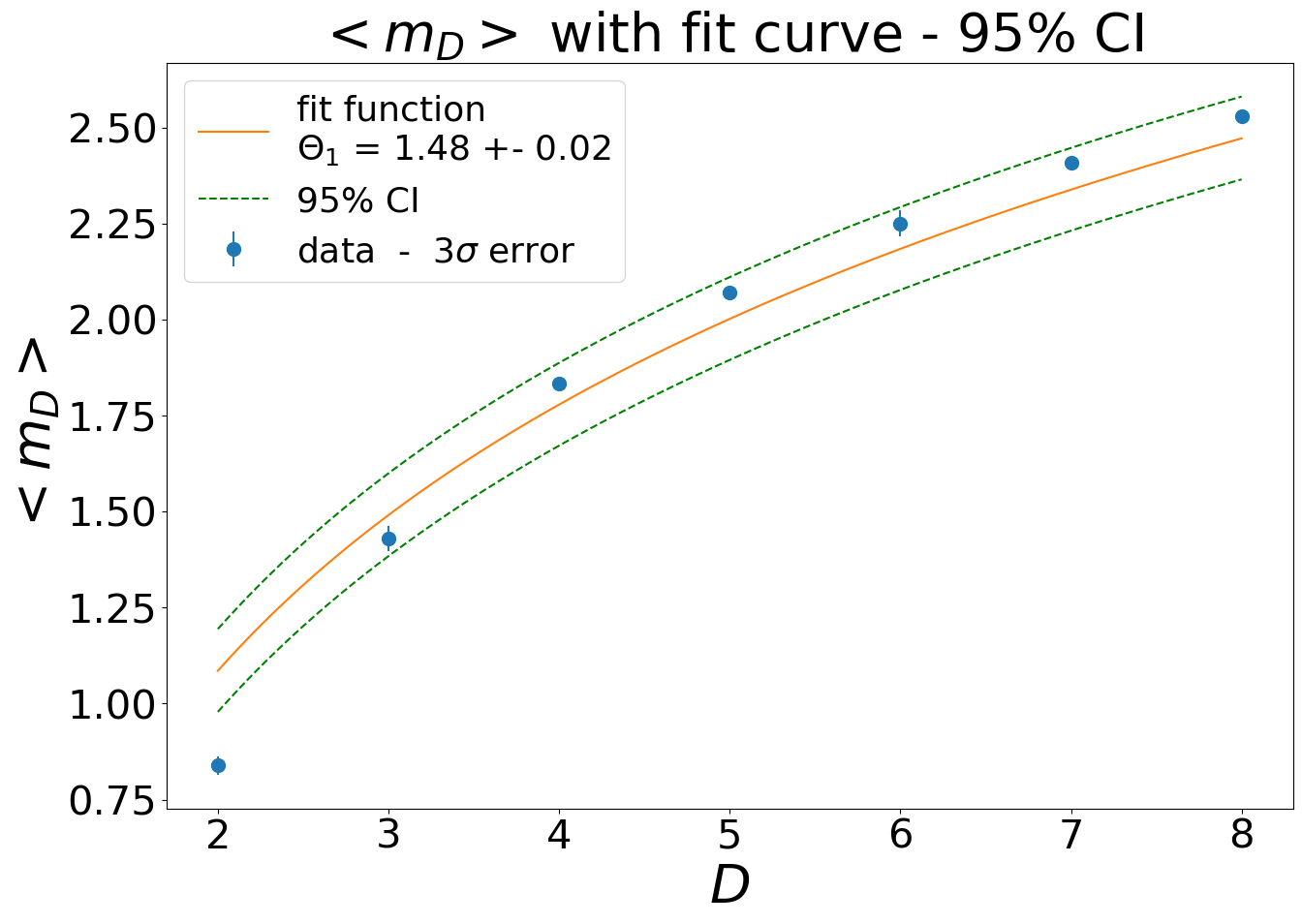}\label{fig:Fit_minimum_N}} $\qquad$ & 
		\subfloat[]{\includegraphics[width=0.45\textwidth]{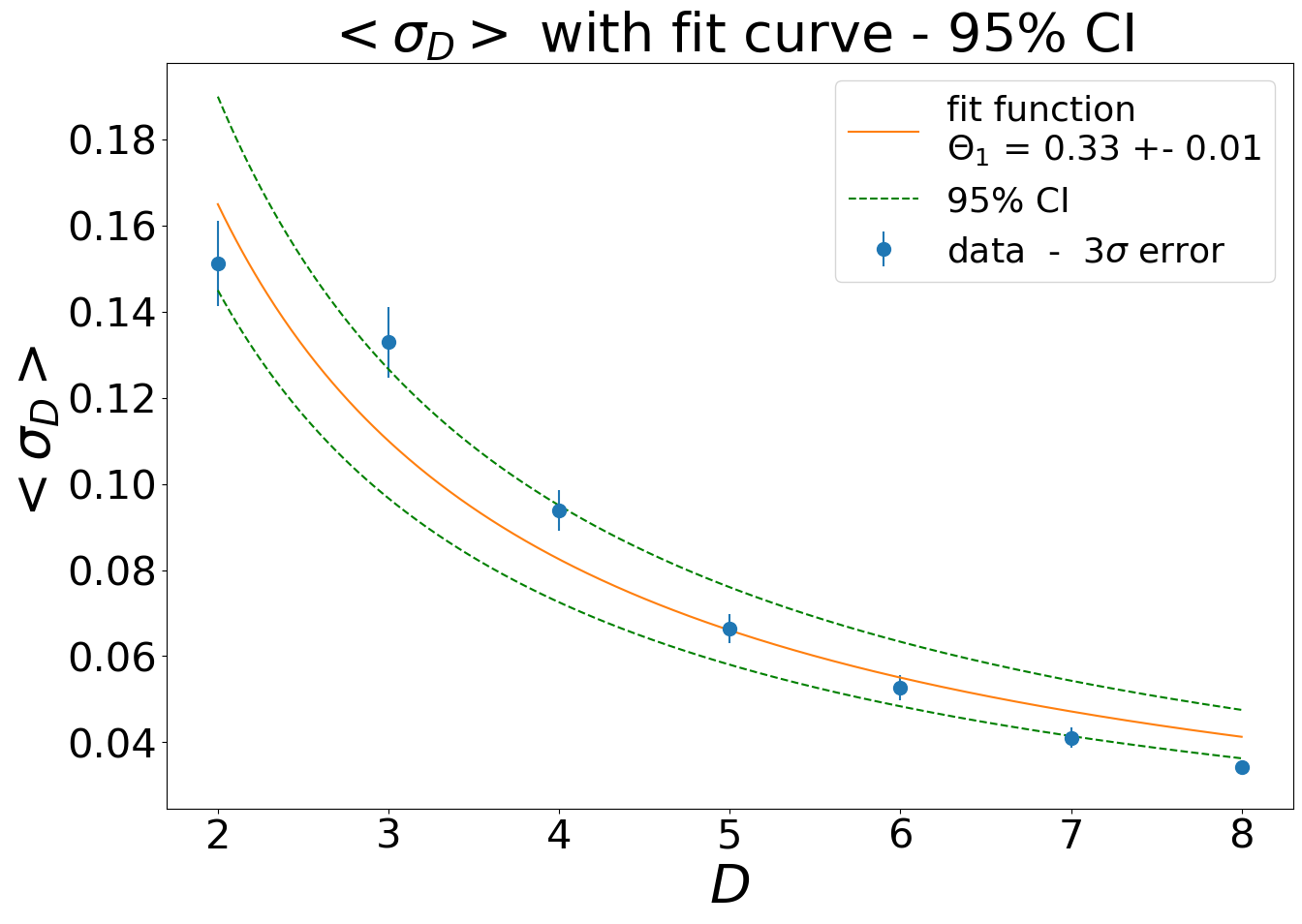}\label{fig:Fit_stdv_N}} \\
		
		% Terza riga con figura + tabella
		\subfloat[]{\includegraphics[width=0.45\textwidth]{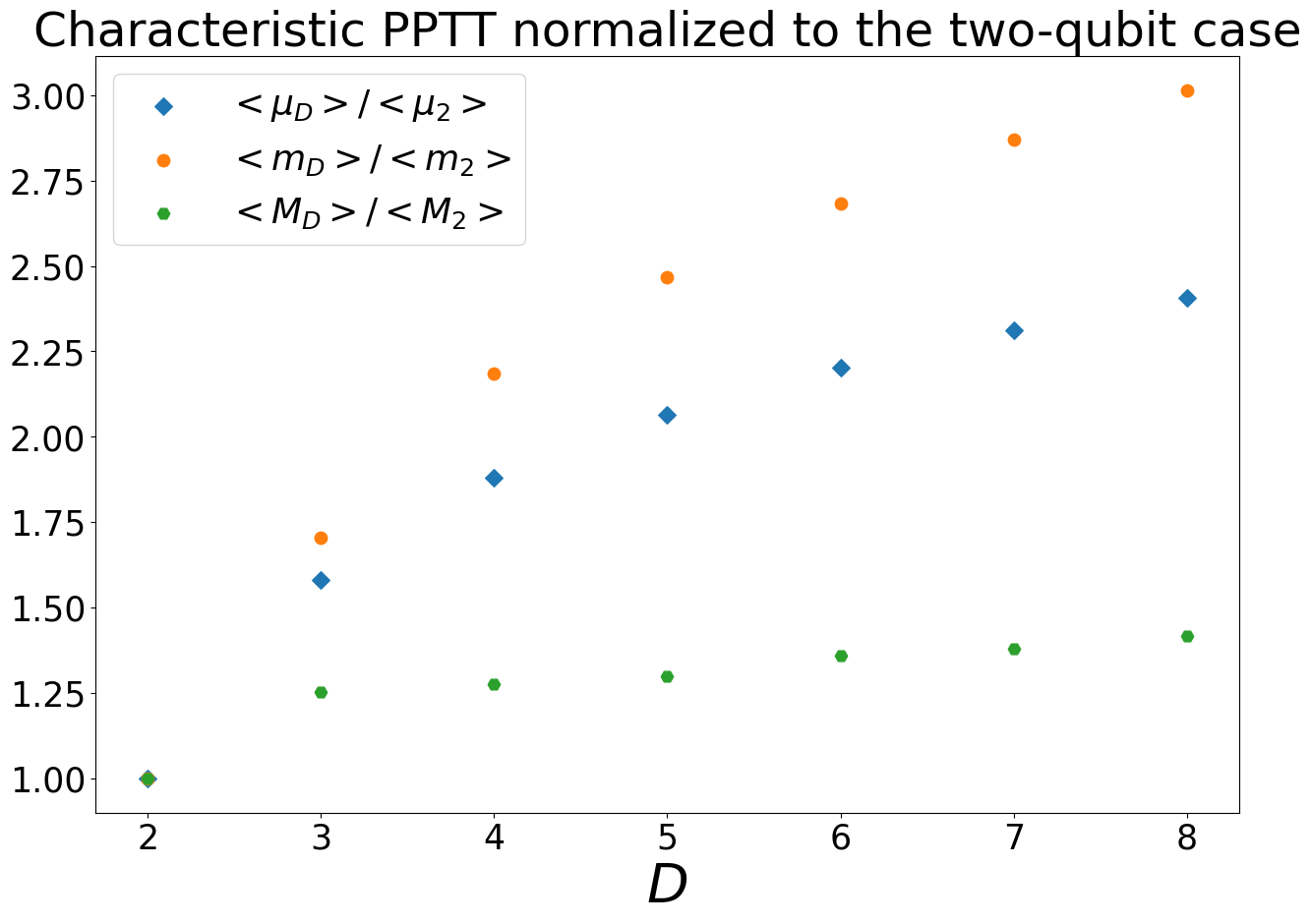}\label{fig:Fit_char_N}} & 
		\begin{minipage}{0.45\textwidth}
			\vspace{-130pt}
			\centering
			\subfloat[\normalsize{Parameter estimates for the characteristic times and standard deviation of the PPTT distributions ${\mathbf P}_{ppt}(x)$ as a function of the dimension $D$.}]{
			\begin{tabular}{ccccc}
			\toprule
			$\chi_D$ & $\theta_{\chi_{D}}$ & Estimate & SE Estim & 95\% CI \\
			\midrule
			$\langle \mu_{D} \rangle$ & $\theta_{\mu_{D}}$ & 1,43 & 0,03 & (1,36; 1,51) \\
			$\langle \mathcal{M}_{D} \rangle$ & $\theta_{\mathcal{M}_{D}}$ & 1,42 & 0,03 & (1,34; 1,50)\\
			$\langle m_{D} \rangle$ & $\theta_{m_{D}}$ & 1,25 & 0,02 & (1,20; 1,29) \\
			$\langle \sigma_{D} \rangle$ & $\theta_{\sigma_{D}}$ & 0,25 & 0,01 & (0,22; 0,27)\\
			\bottomrule
			\end{tabular}
			\label{tab:param_estim_mean_median_min_stdv_D}}
		\end{minipage}
	\end{tabular}
	\caption{Functional dependence of the mean time $\langle \mu_{D} \rangle $ (Panel~\ref{fig:Fit_mean_N}), median time 
		$\langle \mathcal{M}_{D} \rangle $ (Panel~\ref{fig:Fit_median_N}), minimum time $\langle m_{D} \rangle $ (Panel~\ref{fig:Fit_minimum_N}), and 
		standard deviation $\langle \sigma_{N} \rangle $ (Panel~\ref{fig:Fit_stdv_N}) of the empirical PPTT distribution $\mathbf{P}_{ppt}$ with respect to the dimension $D$, obtained from the data of the histograms of Fig.~\ref{fig:PPTT_distrib_increasing_N}. Dashed green lines represent the fit-curves obtained considering a 95\% CI for the fit parameters (See Table~\ref{tab:param_estim_mean_median_min_stdv_D}). Panel~\ref{fig:Fit_char_N}: functional dependence of the median, minimum, and maximum PPT times, normalized to the two-qubit case, with respect to the system dimension $D$.}
	\label{fig:param_estim_mean_median_min_stdv_N}
\end{figure*}

Figure~\ref{fig:param_estim_mean_median_min_stdv_N} illustrates the functional dependence of the mean time ($\mu_{D}$), 
median time  ($\mathcal{M}_{D}$), minimum time ($m_D$), and standard deviation ($\sigma_D$) on $D$, as extrapolated from the data in Fig.~\ref{fig:PPTT_distrib_increasing_N}. The figure also includes the chosen fit functions (discussed below) along with the corresponding fit curves within a $95\%$ confidence interval.
To quantify the uncertainty in the data points, we employed the Bootstrap method on the original sample to estimate the characteristic times shown in the figure. This analysis provides a quantitative confirmation of the trends observed in Fig.~\ref{fig:PPTT_distrib_increasing_N}, revealing a logarithmic growth in the mean, median, and minimum times as $D$ increases. Accordingly, we adopt a fit ansatz of the form~$\log_2{(\theta D)}$ for these quantities.
In contrast, the standard deviation follows a different trend with $D$, for which we use the ansatz  ${\theta}/{D}$, confirming that as the system dimension increases, the PPTT distributions become more sharply peaked. The computed values of the parameters characterizing these fit functions are reported in Table~\ref{tab:param_estim_mean_median_min_stdv_D}. Additionally, the standard error of the estimate (SE Estimate) provides insight into the precision of the computed parameters -- smaller values indicate higher precision.

\subsection{An ansatz for the PPTT distributions for large $D$ values} \label{sec:ansatz} 

In this section we formulate an ansatz for the functional dependence of the PPTT distribution of the system
at large $D$, based on the empirical evidence reported in the previous section. 
To begin with, it is useful to note  that by absorbing the function $\xi_D$ from Eq.~(\ref{constraints})
which determines the normalization of the measure $d\mu^{(\xi_D,r_D)}_D(K)$
 into the parameter $\gamma$ of Eq.~(\ref{defLL}),
 we can leverage the scaling identities~(\ref{IMPO}) to establish relationships between  the PPTT distributions generated
 by different samplings strategies. 
For instance, let 
 $P_{ppt}(\tau)$, $\bar{P}_{ppt}(T)$ denote  the PPTT distribution and its corresponding cumulative distribution obtained by sampling the $(D^2-1)\times (D^2-1)$ Kossakowski matrix  $K$ from the Wishart set with normalization function $\xi_D$. Similarly, let 
$P'_{ppt}(\tau)$ and $\bar{P}'_{ppt}(T)$ represent the distributions   
obtained using a different  function $\xi_D'$. Then, we can express the relationship between these distributions as follows:

\begin{eqnarray}\label{IMPOnew}  
\begin{cases} 
P'_{ppt}(\tau)  &= \frac{\xi'_D}{\xi_D}  P_{ppt}\left( \frac{\xi'_D}{\xi_D} \tau\right)  \;, \\\\ 
\bar{P}'_{ppt}(T)  &=  \bar{P}_{ppt}\left(\frac{\xi'_D}{\xi_D}T\right) \;,
\end{cases} 
\end{eqnarray} 
which, translated in the rescaled time coordinate $x=\gamma \tau$, can be equivalently expressed as  
\begin{eqnarray}\label{IMPOnew1}  
\begin{cases} 
 &{\mathbf P}'_{ppt}(x)  = \frac{\xi'_D}{\xi_D}   {\mathbf P}_{ppt}\left( \frac{\xi'_D}{\xi_D} x\right)  \;, \\\\ 
&\bar{\mathbf P}'_{ppt}(X)  = \bar{\mathbf P}_{ppt}\left(\frac{\xi'_D}{\xi_D}X\right) \;.
\end{cases} 
\end{eqnarray} 
Similarly, indicating with $\chi'_D$ the mean time  $\mu'_{D}$, the median time $\mathcal{M}'_{D}$, minimum time $m'_{D}$, and the standard deviation $\sigma'_{D}$ of ${\mathbf P}'_{ppt}(x)$, and with  $\chi_D$ the corresponding  values $\mu_{D}$,  $\mathcal{M}_{D}$,  $m_{D}$, and  $\sigma_{D}$ associated with ${\mathbf P}_{ppt}(x)$,
from (\ref{IMPOnew1}) it follows that they are related via the identity 
\begin{eqnarray} \label{meancomparison} 
\chi'_{D} = \frac{\xi_D}{\xi'_D} \chi_{D}\;.
\end{eqnarray} 
Suppose we now take 
$\xi_D$ to be the normalization condition
defined in Eq.~(\ref{canonical})
and 
\begin{eqnarray}
\xi'_D= D \; \mu_D\;,
\end{eqnarray}  with $\mu_D$ the  mean time value
of ${\mathbf P}_{ppt}(x)$, so that $\xi_D/\xi'_D= 1/\mu_D$.
From  (\ref{meancomparison}) it thus follows that the mean value and standard deviation of the distribution 
${\mathbf P}'_{ppt}(x)$ fulfil the identity
\begin{eqnarray} 
\mu'_D= 1 \;, \qquad   \qquad \sigma_D' = \sigma_D/\mu_D\;.   
\end{eqnarray} 
Accordingly if now we take for granted that $\mu_D$ and $\sigma_D$ maintain for large $D$ the functional behaviour observed in 
Fig.~\ref{fig:param_estim_mean_median_min_stdv_N} it follows that $\sigma_D'$ will approach $0$ as
with a scaling of the form $1/(D \ln D)$ centered in $x=\mu_D'=1$, i.e. 
 \begin{eqnarray}
 \sigma_D' \simeq  \left. \frac{\theta_{\text{stdv}}}{D \ln(\theta_{\text{mean}} D)} \right|_{D\rightarrow \infty} \longrightarrow 0 \;,  
 \end{eqnarray} 
 implying that ${\mathbf P}'_{ppt}(x)$ will approach a Dirac-delta 
 \begin{eqnarray} \begin{cases}
 &\left. {\mathbf P}'_{ppt}(x) \right|_{D\rightarrow \infty} \longrightarrow \delta(x-1)\;,\\ \\
 &\left. \bar{\mathbf P}'_{ppt}(x) \right|_{D\rightarrow \infty} \longrightarrow \Theta(x-1)\;,
 \end{cases} 
 \end{eqnarray} 
with $\Theta(\cdot)$ the Heaviside step function.

\begin{figure*}
	\subfloat[][]{\includegraphics[scale=0.23]{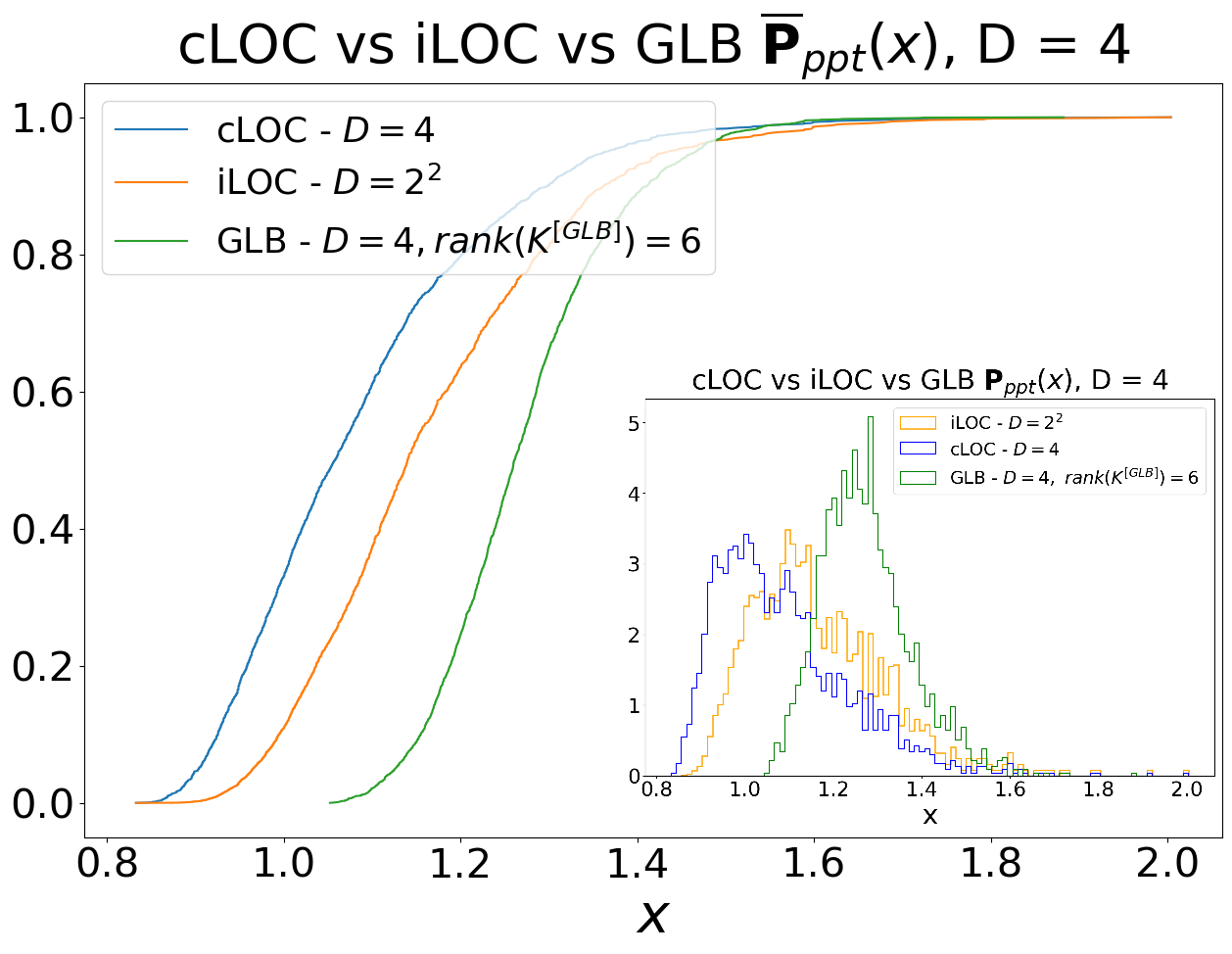}\label{fig:glob_vs_loc_tot_dim_4}} \quad
	\subfloat[][]{\includegraphics[scale=0.23]{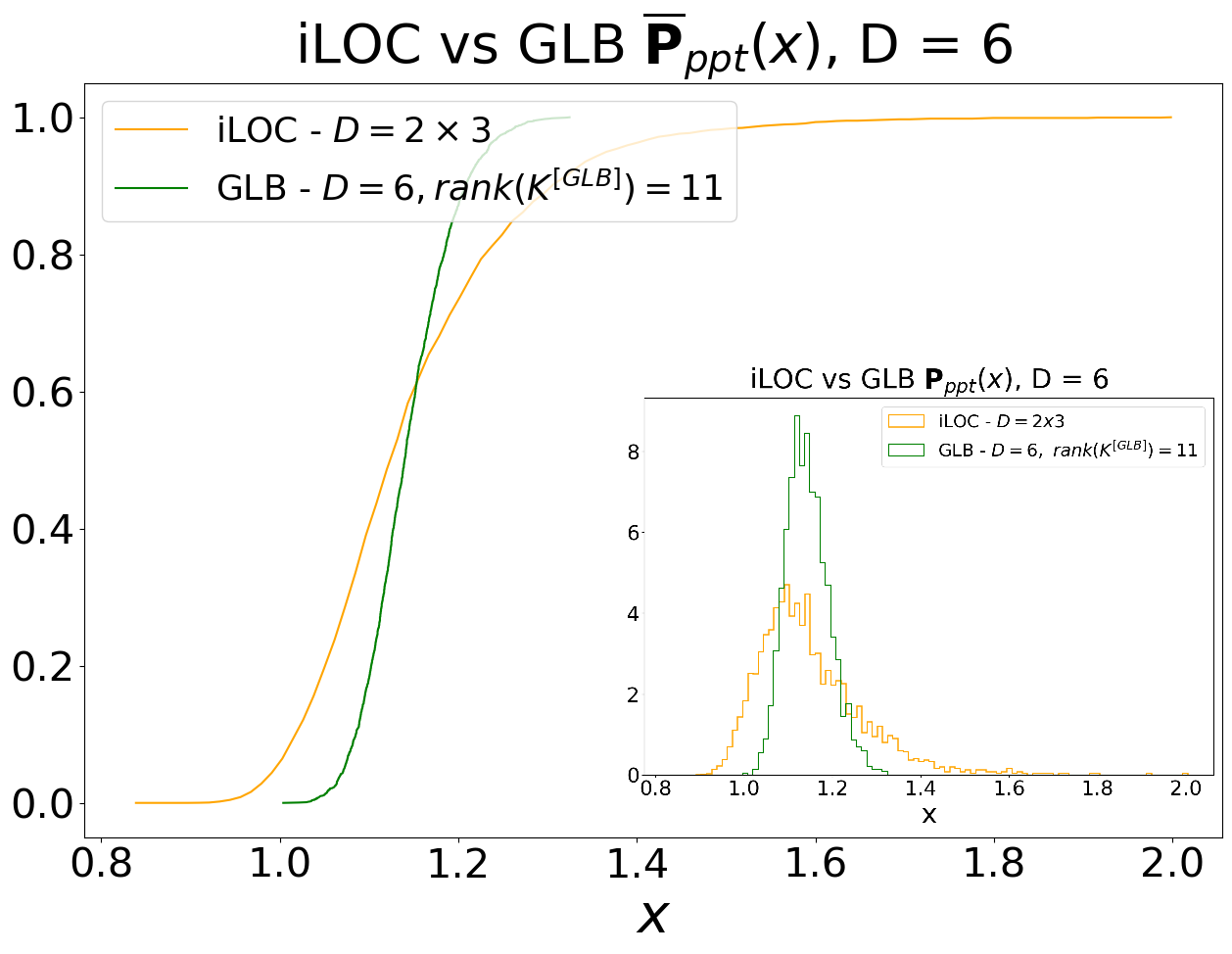}\label{fig:glob_vs_loc_tot_dim_6}} \\
	\subfloat[][]{\includegraphics[scale=0.23]{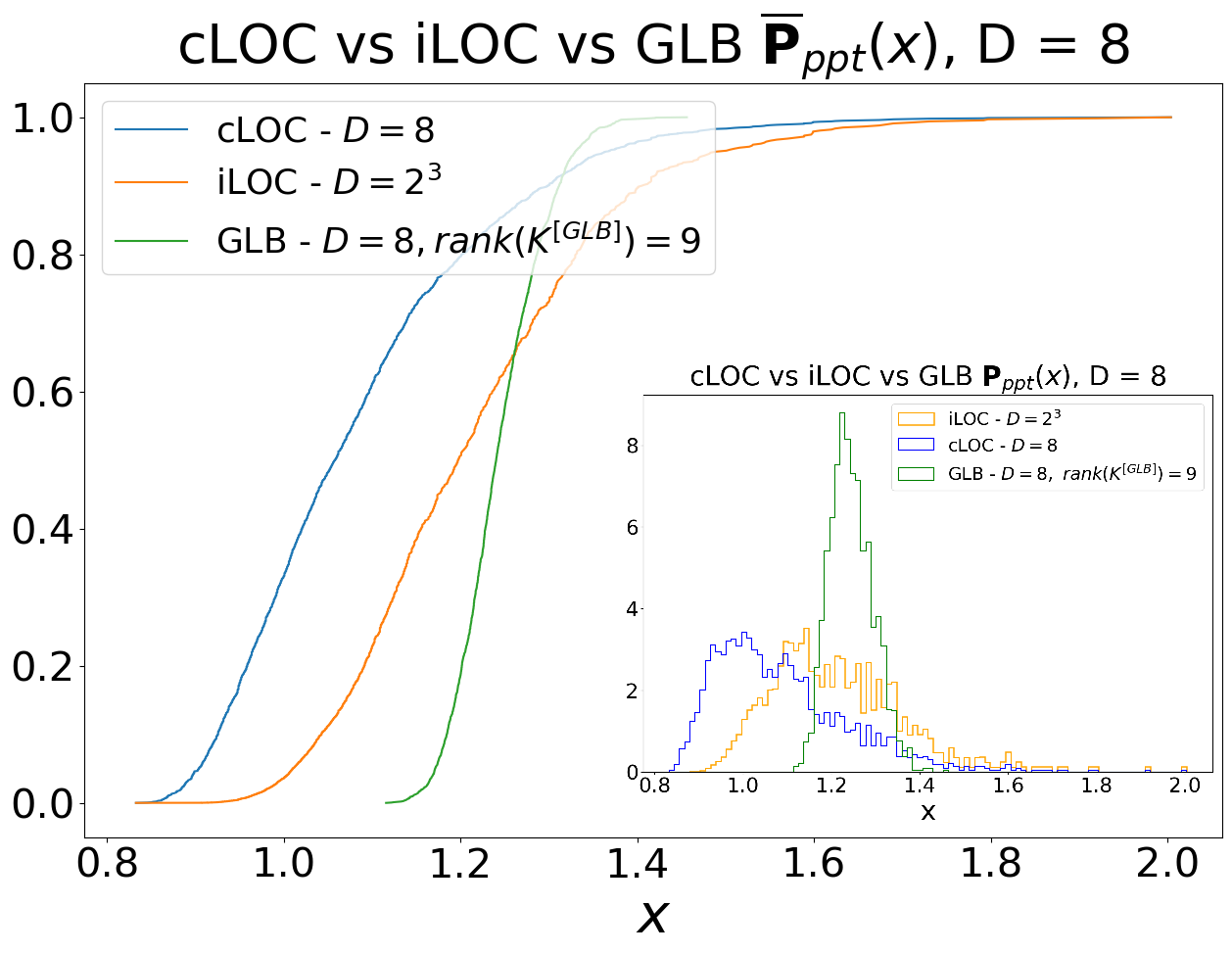}\label{fig:glob_vs_loc_tot_dim_8_rankD9}} \quad
	\subfloat[][]{\includegraphics[scale=0.23]{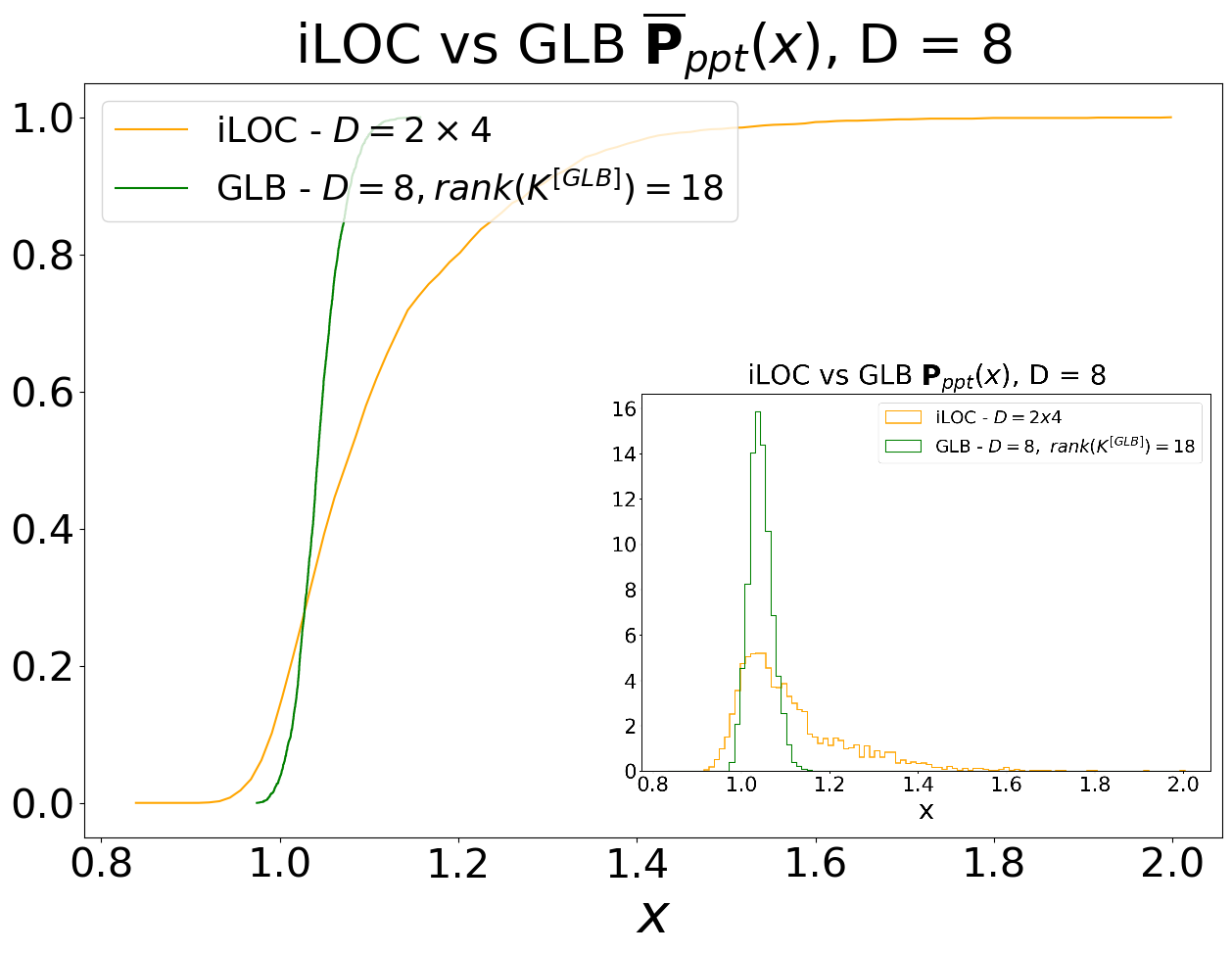}\label{fig:glob_vs_loc_tot_dim_8_rankD18}}
	\caption{Rescaled  PPTT distributions ${\mathbf P}^{[\rm{GLB}]}_{ppt}(x)$ and ${\mathbf P}^{[\rm{iLOC}]}_{ppt}(x)$ (insets) and their cumulative
	counterparts $\bar{\mathbf P}^{[\rm{GLB}]}_{ppt}(x)$ and $\bar{\mathbf P}^{[\rm{iLOC}]}_{ppt}(x)$ (main panels),
	for a composite system $A$  of global dimension $D$ formed by $n$ subsystems of local dimensions $d_1$, $d_2$, $\cdots$, $d_n$ under local and global noise for different subsystems' configurations under same-trace (\ref{sametrace}), same-rank (\ref{samerank}) sampling conditions.
		Panel~\ref{fig:glob_vs_loc_tot_dim_4}: $D=4$, $n=2$, with $d_1=d_2=2$ (isomorphic decomposition),
		which according to Eq.~(\ref{samerank}) implies $r_{d_j=2}= 3$ and $r^{[{\rm GLB}]}_{D=4} = 3+3=6$.
Here we also report the value of the distribution ${\mathbf P}^{[\rm{cLOC}]}_{ppt}(x)$ of the 
		cLOC scenario
		 computed according to Eq.~(\ref{easyrecursiveN1}).
		Panel \ref{fig:glob_vs_loc_tot_dim_6}: $D=6$, $n=2$, with $d_1=3$ and $d_2=2$ (non-isomorphic decomposition), which implies  $r_{d_1=3}= 8$ and $r_{d_2=2}= 3$ (unconstrained rank) for the local subsystems  and  $r^{[{\rm GLB}]}_{D=8} = 8+3=11$ for  $K^{[{\rm GLB}]}$.
		Panel~\ref{fig:glob_vs_loc_tot_dim_8_rankD9}: $D=8$, $n=3$, with $d_1=d_2=d_3=2$  (isomorphic decomposition) which imply $r_{d_j=2}= 3$, and $r^{[{\rm GLB}]}_{D=8} = 3+3+3=9$. 
		 In this case we also report the  distribution ${\mathbf P}^{[\rm{cLOC}]}_{ppt}(x)$.
Panel \ref{fig:glob_vs_loc_tot_dim_8_rankD18}: $D=8$, $n=2$, with $d_1=4$, $d_2=2$ (non-isomorphic decomposition), which implies  $r_{d_1=4}= 15$ and $r_{d_2=2}= 3$ (unconstrained rank), and 
 $\text{rank}(K^{[{\rm GLB}]}) = 18$. 
		All the plots have been realized for purely dissipative GKSL generators ($k=0$). 
		 }
	\label{fig:global_vs_local_noise_all_tot_dim}
\end{figure*}

\section{Correlated Noise and PPTT Distributions}\label{sec:LNvsGN}

This final section examines how different types of correlations influence the PPTT distributions of composite quantum systems. As in previous analyses, we focus on purely dissipative GKLS generators by setting~$k=0$ in Eq.~(\ref{defLL}). However we modify the scaling of the normalization function $\xi_D$ of  Eq.~(\ref{constraints}) and the rank parameter $r_D$ of Eq.~(\ref{constraints2}) from the values we assumed in 
Sec.~\ref{Sect:PPTT_RandomNoise}, to prevent biases arising from the different sampling techniques used for various noise models -- see details below.

\subsection{Composite quantum system} 
Assume that, as depicted in Fig.~\ref{figura1}, our $D$-dimensional quantum system $A$ is now
composed of $n$ (potentially heterogeneous) subsystems $a^{(j)}$ with dimensions $d_1, d_2, \dots, d_n$, such that 
\begin{eqnarray}
D = d_1 d_2 \dots d_n\;. \end{eqnarray}  Within this framework, we identify the following scenarios:

{\bf i) Statistically independent local noise sources} [{iLOC}]:
In this case (corresponding to panels b) and c) of the figure), the dissipator governing the Markovian evolution of $A$ is given by the sum of local terms acting independently on the individual subsystems. Accordingly, the dynamical generator of the full system takes the form\begin{equation}\label{locL}  
{\cal D}^{[\rm{LOC}]}_{K_1,\dots, K_n} = {\cal D}_{K_1}^{(1)} + {\cal D}_{K_2}^{(2)} + \dots + {\cal D}_{K_n}^{(n)}\;,  
\end{equation}  
where, for each $j=1,\cdots, n$, the term ${\cal D}^{(j)}_{K_j}$ denotes
 the local dissipator acting on subsystem $a^{(j)}$, associated with a $(d_j^2-1)
\times (d_j^2-1)$ Kossakowski matrix $K_j$. 
As shown in Appendix~\ref{sec:norm},  when ${\cal D}^{[\rm{LOC}]}_{K_1,\dots, K_n}$ is expressed in the canonical form~(\ref{eq:Dissipator_Kmn}) with respect to the orthonormal basis $\{\hat{F}_{m}\}_{m=1,\cdots,D^2 -1}$ of the joint system $A$, 
the corresponding Kossakowski matrix $K^{[\mathrm{LOC}]}$  is a $(D^2 - 1) \times (D^2 - 1)$ block-diagonal matrix given by
\begin{eqnarray}{K}^{[\rm{LOC}]}= \label{block} 
\bigoplus_{j=1}^{n} 
 \frac{D}{d_j} K_j\;,
\end{eqnarray} 
where each block corresponds to one of the local Kossakowski matrices, rescaled by the ratio $D/d_j$.
In the iLOC  scenario, we generate the local contributions in Eq.~(\ref{locL}) independently, according to the product measure
\begin{eqnarray}\label{mulocL}  
&& d\mu^{[\rm{iLOC}]}_D(K_1,\dots, K_n) \\ \nonumber 
&&\qquad := d\mu^{(\xi_{d_1},r_{d_1})}_{d_1}({K_1}) \cdots d\mu^{(\xi_{d_n},r_{d_n})}_{d_n}({K_n})\;,  
\end{eqnarray}  
where $d\mu^{(\xi_{d_j},r_{d_j})}_{d_j}({K_j})$ denotes the probability measure used to sample $K_j$ from the set of Wishart $(d_j^2 - 1) \times (d_j^2 - 1)$ matrices with 
 constraints ~(\ref{constraints}), (\ref{constraints2}) assigned by properly selected functions
 $\xi_{d_j}$ and $r_{d_j}$.

{\bf ii) Correlated local noise sources} [{cLOC}]:
In this scenario (corresponding to panel d) of  Fig.~\ref{figura1}), the GKLS generators of $A$ are still expressed as a sum of local terms as in Eq.~(\ref{locL}). However, unlike the iLOC case, the noise generators of individual subsystems are now strongly correlated. Specifically, once a single generator (e.g., $K_1$, acting on $a^{(1)}$) is assigned, all other local GKLS generators in Eq.~(\ref{locL}) are uniquely determined by predefined functional mappings:  
\begin{equation}  
K_j = {\cal F}_j(K_1), \quad \forall j \geq 2.  
\end{equation}  
Consequently, the measure in Eq.~(\ref{mulocL}) is modified as  
\begin{eqnarray}\label{correlocL}  
&&d\mu^{[\rm{cLOC}]}_D(K_1,\dots, K_n)  \\ \nonumber  
&&\qquad = d\mu^{(\xi_{d_1},r_{d_1})}_{d_1}({K_1}) \prod_{j=2}^n \delta(K_j - {\cal F}_j(K_1)) \;.  
\end{eqnarray}

{\bf iii) Global noise sources} [{GLB}]: 
In this case (corresponding to panel a) of  Fig.~\ref{figura1}), the dissipator governing $A$ includes non-local contributions that induce couplings among the subsystems. The  sampling of the associated Kossakowski matrix $K^{[\rm{GLB}]}$ is performed
 using the usual measure over the set of  Wishart $(D^2-1)\times(D^2-1)$ matrices with constraints ~(\ref{constraints}), (\ref{constraints2}) assigned by properly selected functions
 $\xi_D^{[\rm{GLB}]}$ and $r_D^{[\rm{GLB}]}$.

\subsection{Homogenous sampling conditions} \label{sec:comp} 
For the two local scenarios (iLOC and cLOC), each instance of the selected generator defines an evolution of the system $A$ as a product of local quantum channels:  
\begin{equation} \label{q-chloc}  
\Phi_{(0,t)} = \Phi^{(1)}_{(0,t)} \otimes \cdots \otimes \Phi^{(n)}_{(0,t)}\;,  
\end{equation}  
where $\Phi^{(j)}_{(0,t)} := e^{\gamma (t - t_0) {\cal D}_{K_j}^{(j)}}$ acts locally on subsystem $j$.  
As shown in Ref.~\cite{CGDP1} in the iLOC case, the cumulative distribution of the PPT time follows a simple product rule:  
\begin{equation} \label{easyrecursiveN}  
\bar{\mathbf P}^{[\rm{iLOC}]}_{ppt}(x) = \bar{\mathbf P}^{(1)}_{ppt}(x) \cdots \bar{\mathbf P}^{(n)}_{ppt}(x)\;.  
\end{equation}  
Differentiating this, we obtain  
\begin{eqnarray} \label{easyrecursiveNp}  
{\mathbf P}^{[\rm{iLOC}]}_{ppt}(x) &=& \frac{d \bar{\mathbf P}^{[\rm{iLOC}]}_{ppt}(x)}{d x} \\  
\nonumber &=&  \bar{\mathbf P}^{[\rm{iLOC}]}_{ppt}(x) \sum_{j=1}^n \frac{{\mathbf P}^{(j)}_{ppt}(x)}{ \bar{\mathbf P}^{(j)}_{ppt}(x)} \;.  
\end{eqnarray}  
The cLOC case is more intricate, as the correlations between local generators significantly impact the sampling process. However, in the special case where all subsystems are isomorphic and subject to the same noise, the system remains PPT as soon as one of the individual maps $\Phi^{(j)}_{(0,t)}$ (e.g., $\Phi^{(1)}_{(0,t)}$) becomes PPT. Thus, we obtain  
\begin{eqnarray} \label{easyrecursiveN}  
{\mathbf P}^{[\rm{cLOC}]}_{ppt}(x) &=&  {\mathbf P}^{(1)}_{ppt}(x)  \;, \\  
\bar{\mathbf P}^{[\rm{cLOC}]}_{ppt}(x) &=&   \bar{\mathbf P}^{(1)}_{ppt}(x)  \;.  \label{easyrecursiveN1} 
\end{eqnarray}  
For the GLB scenario, the PPTT distribution ${\mathbf P}^{[\rm{GLB}]}_{ppt}(x)$ is obtained instead  by sampling the generator on $A$ without imposing restrictions on correlations.  

A fair comparison of the PPTT distributions across the iLOC, cLOC, and GLB scenarios 
requires ensuring that the different sampling techniques used for the associated Kossakowski matrices do not introduce biases.
One possible approach would be to generate the samples for ${\mathbf P}^{[\rm{iLOC}]}_{ppt}(x)$ and ${\mathbf P}^{[\rm{cLOC}]}_{ppt}(x)$ by post-selecting from the $K$ matrices obtained during the sampling of ${\mathbf P}^{[\rm{GLB}]}_{ppt}(x)$, retaining only those whose associated dissipator exhibits the proper local structure~(\ref{locL}). However, this method is highly impractical, as the occurrence of such post-selected events would be exceedingly rare, making it nearly impossible to build a statistically significant dataset for characterizing the local scenarios.
To address this issue, we act on the normalization condition~(\ref{constraints}), which determines the scaling parameter of the Kossakowski matrices,
and on the rank parameter which via Eq.~(\ref{constraints2}) fixes the maximum number of independent noise channel associated with the Lindblad generator. 
In particular from Eq.~(\ref{block}) we observe that the trace and the rank of the matrix ${K}^{[\rm{LOC}]}$ which defines the local dissipator~(\ref{locL}) fulfil the relations
\begin{equation}
\begin{cases}\mbox{Tr}[{K}^{[\rm{LOC}]}]= \label{block11} 
\sum_{j=1}^{n} 
 \frac{D}{d_j} \mbox{Tr}[K_j]=\sum_{j=1}^{n} 
 \frac{D}{d_j} \xi_{d_j}\;, \\\\
 \mbox{rank}[{K}^{[\rm{LOC}]}]= 
\sum_{j=1}^{n} 
\mbox{rank}[K_j] \leq \sum_{j=1}^{n} 
r_{d_j} \;. 
\end{cases}
\end{equation} 
Accordingly we can force the trace of ${K}^{[\rm{LOC}]}$ to be equal to the trace of the Kossakowski matrix of the global scenario by
adopting 
 a super-linear dependence of the form ${\rm x} \log_2 {\rm x}$ on system size for the normalization condition~(\ref{constraints}). Specifically, we set
\begin{equation} \label{xiJ}  
\xi_{d_j} = d_j \log_2 d_j  \;,
\end{equation}  
when sampling subsystems $a^{(j)}$ in the iLOC and cLOC cases, and  
\begin{equation}  
\mbox{Tr}[ {K}^{[\rm{GLB}]}] = \xi^{[\rm{GLB}]}_D = D \log_2 D  \;,
\end{equation}  
when sampling the Kossakowski matrix $K^{[\rm{GLB}]}$ of the GLB scenario.  
From Eq.~(\ref{block11}) it hence follows that at the level of the total system $A$, the resulting Kossakowski matrices maintain a consistent trace value across different sampling methods, i.e.
\begin{equation}\label{sametrace}
\mbox{Tr}[ {K}^{[\rm{LOC}]}] =  \sum_{j=1}^n  {D} \log_2{d_j} =\mbox{Tr}[ {K}^{[\rm{GLB}]}] \;,
\end{equation} 
thereby ensuring 
that the different scenarios are characterized by the same arithmetic mean value of the dissipative rates.
To compensate the difference in the rank instead we  force the parameter $r^{[\rm{GLB}]}_D $ which 
charaterizes the sampling of the global scenario to match the upper bound $\sum_{j=1}^{n} 
r_{d_j}$ on $\mbox{rank}[{K}^{[\rm{LOC}]}]$ induced by the  local rank constraints $r_{d_j}$. 
In particular fixing the  latter at their maximum values
 we take
\begin{eqnarray}
\begin{cases}r_{d_j}=d_j^2-1\;,  \\ \\ 
r^{[\rm{GLB}]}_D = \sum_{j=1}^{n} r_{d_j}= \sum_{j=1}^{n} (d_j^2-1)\;.\label{samerank} \end{cases} 
\end{eqnarray}

\subsection{Numerical results} \label{sec:numres} 

In our simulations, we consider different system configurations based on the dimensions of the Hilbert spaces involved:

\begin{itemize}

\item $D=4$: We  compare the GBL scenario with the iLOC and cLOC scenarios for $n=2$ subsystems, each with local dimension $d_1= d_2=2$. 

\item $D=6$: The GBL scenario is compared with an iLOC configuration consisting of $n=2$ non-isotropic subsystems of dimensions $d_1=3$ and $d_2=2$. 

\item $D=8$: Two alternative iLOC scenarios are considered -- one with $n=2$ subsystems of local dimensions $d_1=4$ and $d_2=2$, 
and another with $n=3$ subsystems where $d_1=d_2=d_3=2$. 
For the latter (isotropic) configuration, we also analyze the associated cLOC scenario.
\end{itemize}

\begin{figure*}
    \subfloat[][]{\includegraphics[scale=0.23]{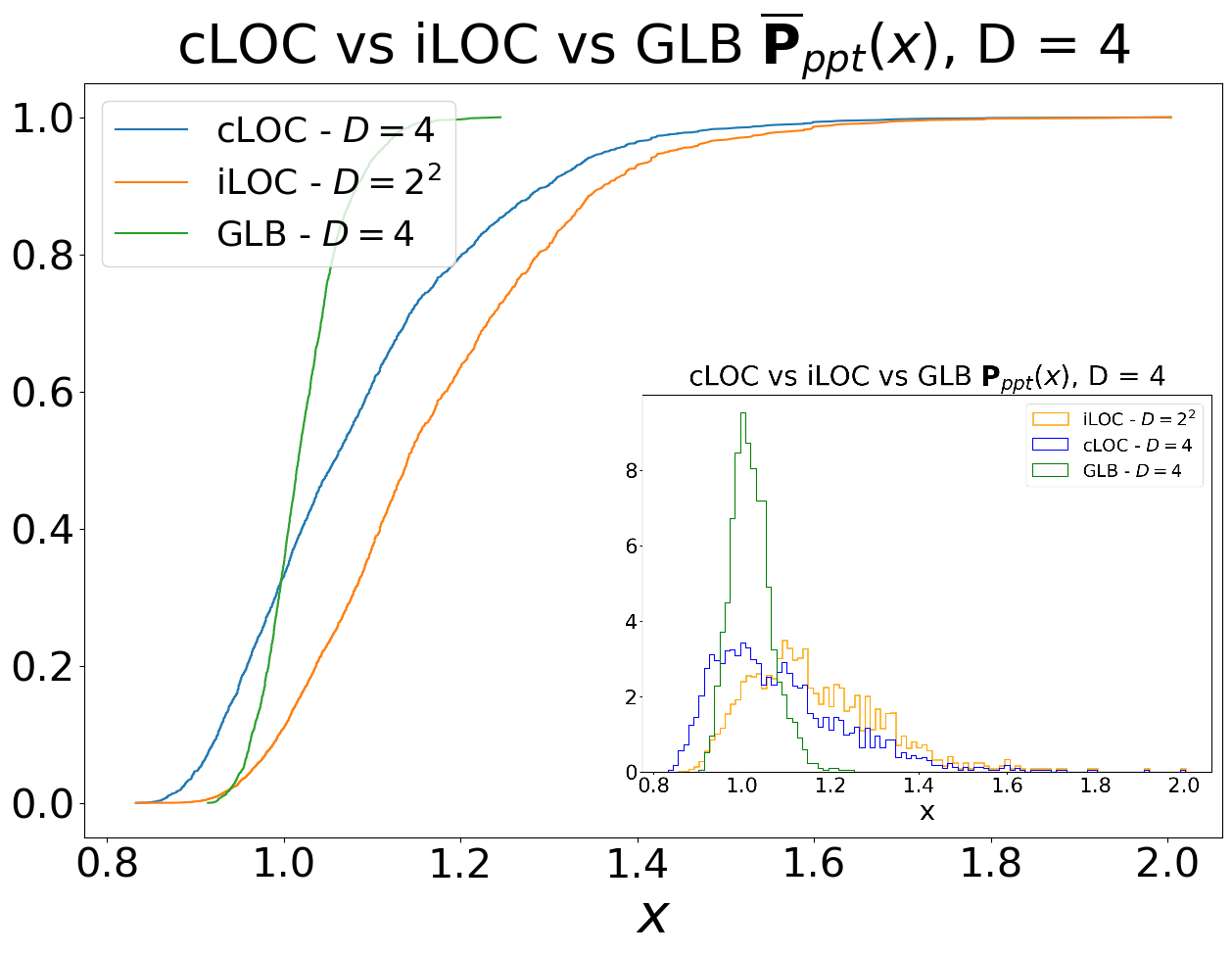}\label{fig:glob_vs_loc_tot_dim_4_nosmallrank}} \quad
    \subfloat[][]{\includegraphics[scale=0.23]{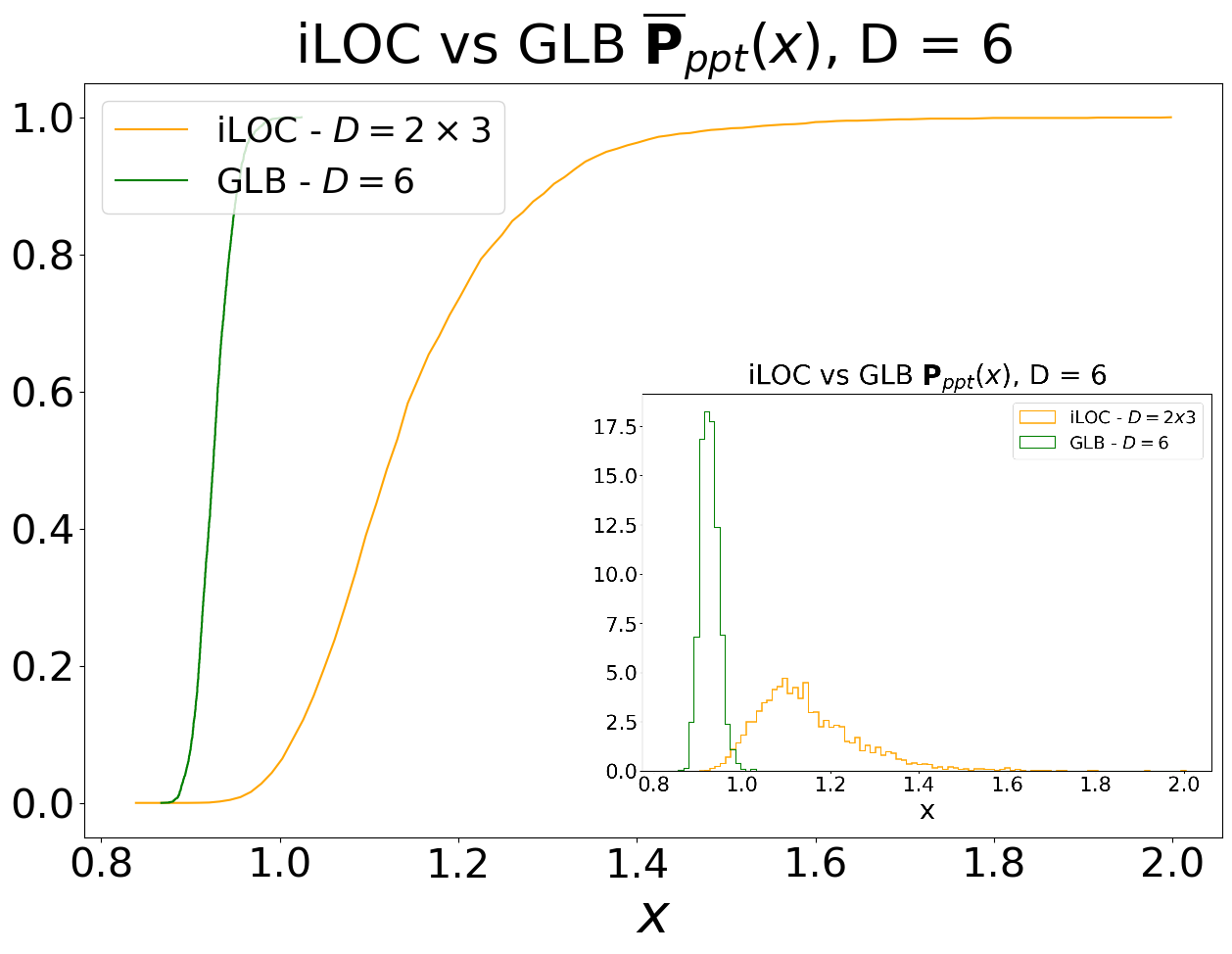}\label{fig:glob_vs_loc_tot_dim_6_nosmallrank}} \\
    \subfloat[][]{\includegraphics[scale=0.23]{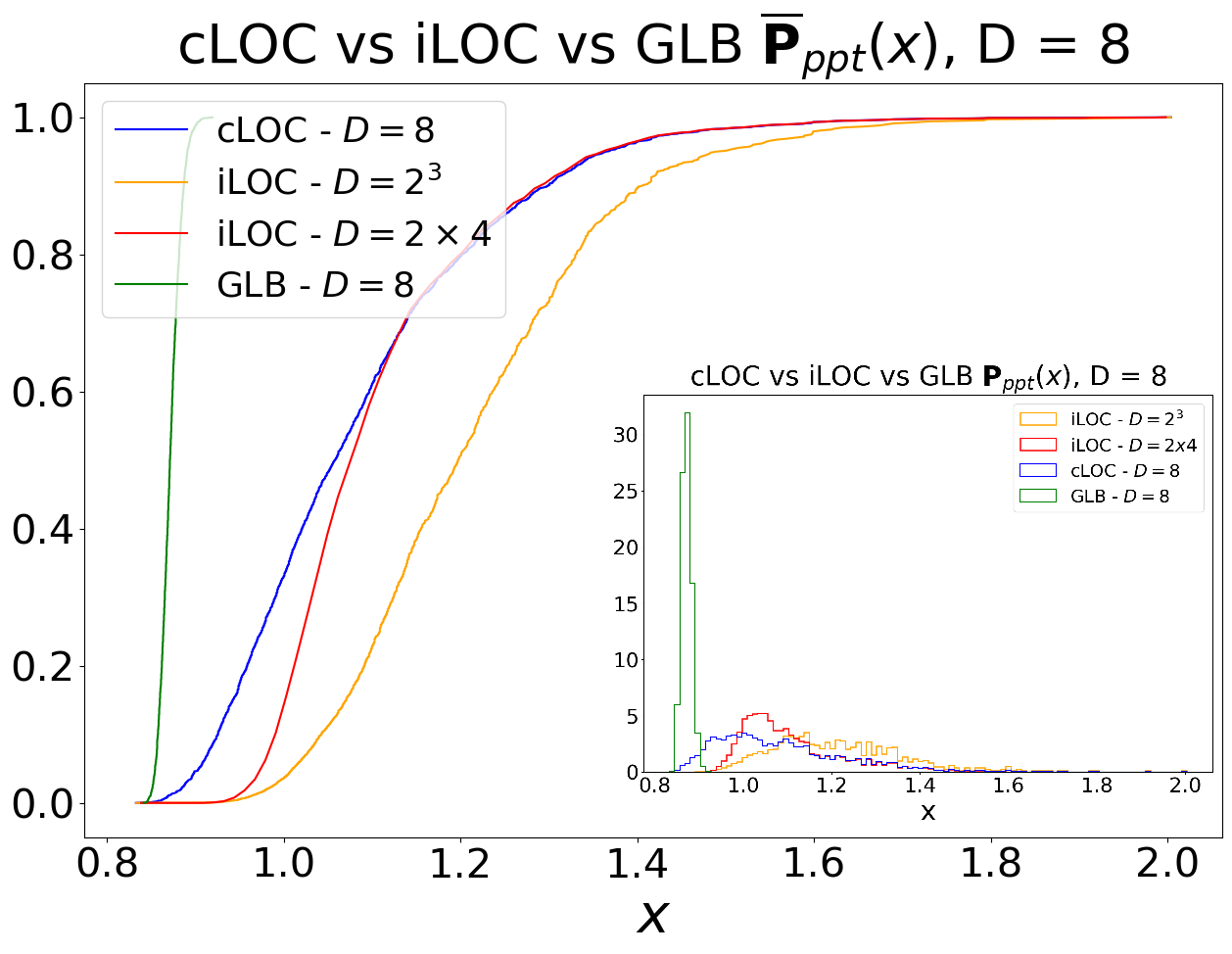}\label{fig:glob_vs_loc_tot_dim_8_nosmallrank}}
    \caption{
Rescaled  PPTT distributions ${\mathbf P}^{[\rm{GLB}]}_{ppt}(x)$ and ${\mathbf P}^{[\rm{iLOC}]}_{ppt}(x)$ (insets) and their cumulative
	counterparts $\bar{\mathbf P}^{[\rm{GLB}]}_{ppt}(x)$ and $\bar{\mathbf P}^{[\rm{iLOC}]}_{ppt}(x)$ (main panels),    
	for a composite system $A$  of global dimension $D$ formed by $n$ subsystems of local dimensions $d_1$, $d_2$, $\cdots$, $d_n$ under local and global noise for different subsystems' configurations under same-trace (\ref{sametrace}) conditions. At variance with Fig.~\ref{fig:global_vs_local_noise_all_tot_dim}
here we do not impose  the same-rank (\ref{samerank}) conditions and assume instead maximum rank values for both the local ($r_{d_j}=d^2_j-1$) and the global
sampling ($r_D^{[\rm{GLB}]}=D^2-1$).  
    Panel a):  $D=4$, $n=2$, with $d_1=d_2=2$ (isomorphic decomposition);
    Panel b): $D=6$, $n=2$, with $d_1=3$ and $d_2=2$ (non-isomorphic decomposition); 
     Panel c): $D=8$
     with two different system decompositions. The first assume $A$ to be formed by three homogeneous subsystems  of dimension $2$, the second instead
     assume $A$ to be formed by two subsystems,  a large one of dimension $d_1=4$   and a  small one of dimension $d_2=2$. For the former configuration we also report the corresponding ${\mathbf P}^{[\rm{cLOC}]}_{ppt}(x)$.
    All the plots have been realized for purely dissipative GKSL generators ($k=0$). } 
    \label{fig:global_vs_local_noise_all_tot_dim1}
\end{figure*}

In Fig.~\ref{fig:global_vs_local_noise_all_tot_dim}, we compare the rescaled distributions
${\mathbf P}^{[\rm{GLB}]}_{ppt}(x)$, ${\mathbf P}^{[\rm{iLOC}]}_{ppt}(x)$,  and, for isotropic configurations, ${\mathbf P}^{[\rm{cLOC}]}_{ppt}(x)$, obtained imposing the
same-trace (\ref{sametrace}) and same-rank (\ref{samerank}) sampling conditions. 
A first observation is that, by construction, the cLOC scenario consistently yields higher cumulative distribution values than the corresponding iLOC scenario. This is expected as from Eqs.~(\ref{easyrecursiveN}) and~(\ref{easyrecursiveN1}) we have $\bar{\mathbf P}^{[\rm{cLOC}]}_{ppt}(x)\geq \bar{\mathbf P}^{[\rm{iLOC}]}_{ppt}(x)$ 
for all $x$.
More importantly, and clearly visible in the plots, the PPTT distributions for the iLOC scenario consistently assign higher probability to smaller PPTTs compared to those of the GLB scenario. This indicates that, under local noise, distillable entanglement tends to decay more rapidly. This difference is particularly pronounced in the isotropic decompositions 
 (Panels~\ref{fig:glob_vs_loc_tot_dim_4} and~\ref{fig:glob_vs_loc_tot_dim_8_rankD9}).
 We emphasize that this behavior crucially depends on the enforcement of homogeneity conditions in the sampling methods. To illustrate this, Fig.~\ref{fig:global_vs_local_noise_all_tot_dim1} presents the rescaled PPTT distributions obtained when only the same-trace condition (\ref{sametrace}) is imposed, while the same-rank condition (\ref{samerank}) is relaxed.
 In this case, although both the GLB and iLOC scenarios use dissipators with the same arithmetic mean of dissipative rates, the higher rank of the Kossakowski matrices in the GLB configuration leads to  ${\mathbf P}^{[\rm{GLB}]}_{ppt}(x)$ assigning greater probability to smaller PPTTs compared to the iLOC configuration.

\section{Conclusions}\label{Sect:Conclusions}

In this work, we have presented a statistical framework to analyze the degradation of entanglement in composite quantum systems subjected to Markovian noise. By evaluating ensembles of GKSL generators and computing the Positive Partial Transpose Time (PPTT), we provided a detailed comparison of how global and local noise models affect entanglement persistence. Our results show that systems exposed to global noise, even under similar average dissipative rates and channel ranks, consistently maintain entanglement for longer durations than those subjected to independent local noise. This highlights the detrimental impact of uncorrelated local dissipation on entanglement resilience.

We demonstrated that the Cao-Lu integration method~\cite{Cao_Lu}  enables efficient and accurate computation of PPTT distributions for systems with dimensions up to $D=8$, significantly reducing computational complexity without compromising accuracy. The resulting empirical distributions suggest a logarithmic growth of characteristic PPTT times with increasing system size, alongside a concurrent narrowing of the distribution, indicating increasingly predictable entanglement lifetimes in larger systems.

Furthermore, our analysis of different noise correlation structures showed that internal configuration -- such as isotropy, subsystem size disparity, and noise correlation -- can critically influence entanglement decay. Notably, even under homogeneous sampling conditions, locally applied noise models yielded shorter PPTTs than global noise models, with correlated local noise scenarios faring slightly better than fully independent ones.

These findings are particularly relevant for quantum technologies that rely on distributed entanglement, such as quantum memories and communication protocols. They underscore the importance of engineering noise environments and system architectures that either exploit global interactions or mitigate the fragmentation effects of independent local dissipation.

Finally, our work sets the stage for future investigations into entanglement dynamics under non-Markovian or structured noise models, as well as the extension to higher-dimensional or multipartite systems. Further exploration into the optimization of numerical techniques, including adaptive step-size control or machine learning-assisted sampling, may also enhance the practical applicability of our statistical approach.

NC and VG acknowledge financial support by MUR (Ministero dell’ Universit\'a e della
Ricerca) through the PNRR MUR project PE0000023-NQSTI. SS acknowledges financial support from National Centre for HPC, Big Data and Quantum Com-
puting (Spoke 10, CN00000013). GDP has been supported by the HPC Italian National Centre for HPC, Big Data and Quantum Computing - Proposal code CN00000013 - CUP J33C22001170001 and by the Italian Extended Partnership PE01 - FAIR Future Artificial Intelligence Research - Proposal code PE00000013 - CUP J33C22002830006 under the MUR National Recovery and Resilience Plan funded by the European Union - NextGenerationEU. Funded by the European Union - NextGenerationEU under the National Recovery and Resilience Plan (PNRR) - Mission 4 Education and research - Component 2 From research to business - Investment 1.1 Notice Prin 2022 - DD N. 104 del 2/2/2022, from title “understanding the LEarning process of QUantum Neural networks (LeQun)”, proposal code 2022WHZ5XH – CUP J53D23003890006. GDP is a member of the “Gruppo Nazionale per la Fisica Matematica (GNFM)” of the “Istituto Nazionale di Alta Matematica “Francesco Severi” (INdAM)”

\bibliographystyle{unsrt}
\bibliography{biblio}

\appendix

\section{Numerical tests} \label{sec:comparison}

\begin{figure*}
    \subfloat[][]{\includegraphics[scale=0.225]{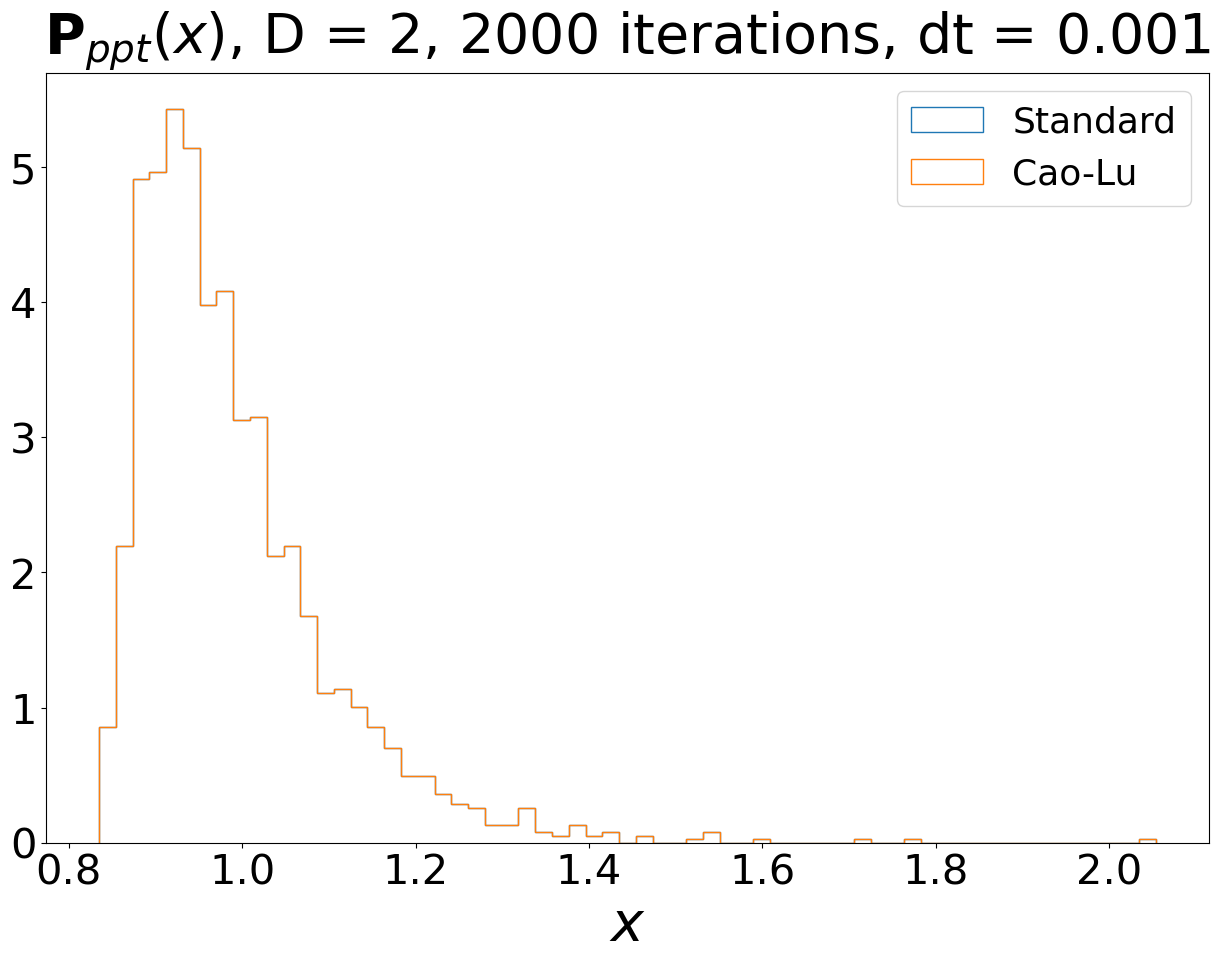}}\label{N=2_dt=0.001} \quad
    \subfloat[][]{\includegraphics[scale=0.225]{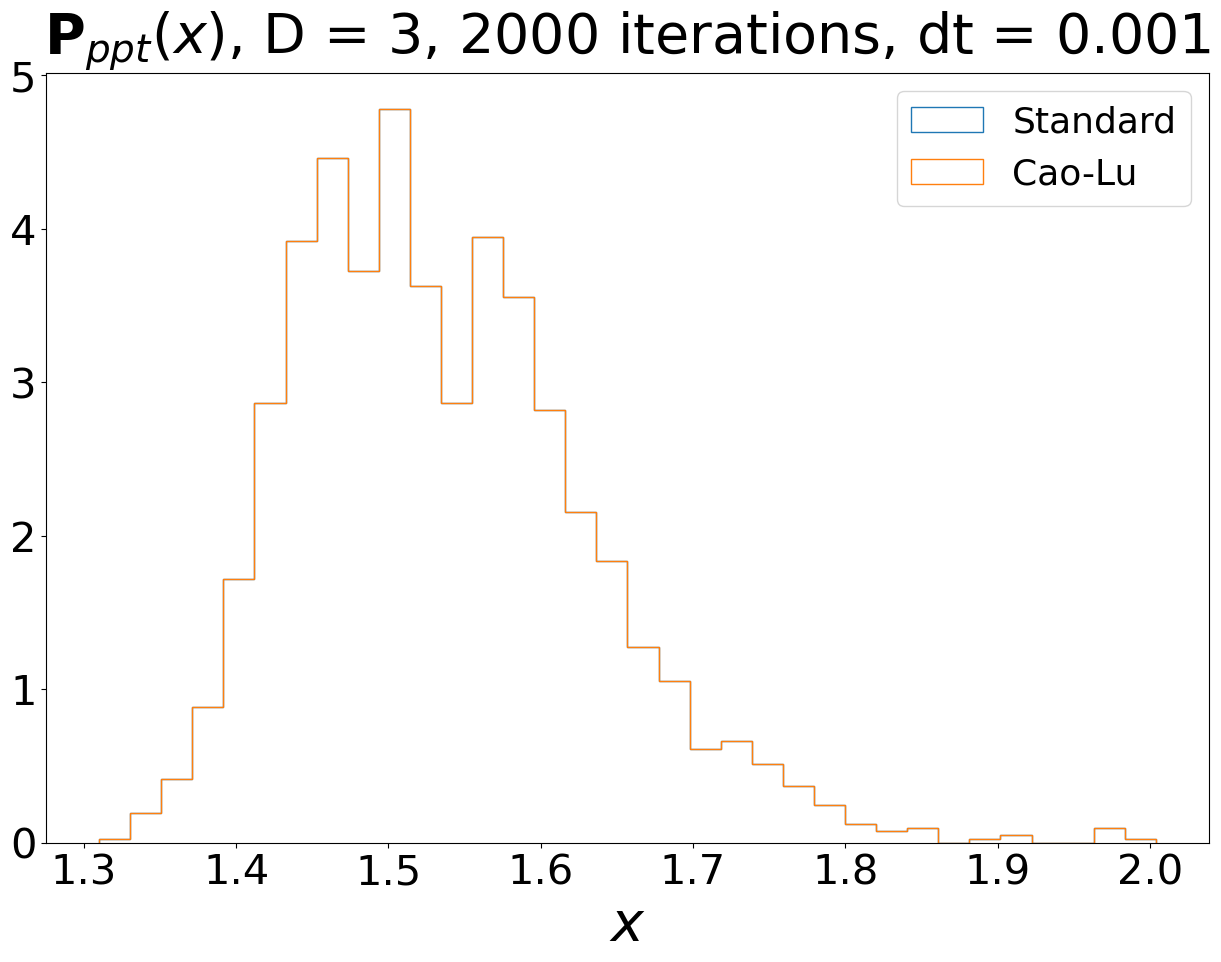}}{\label{N=3_dt=0.001}} \\
    \subfloat[][]{\includegraphics[scale=0.225]{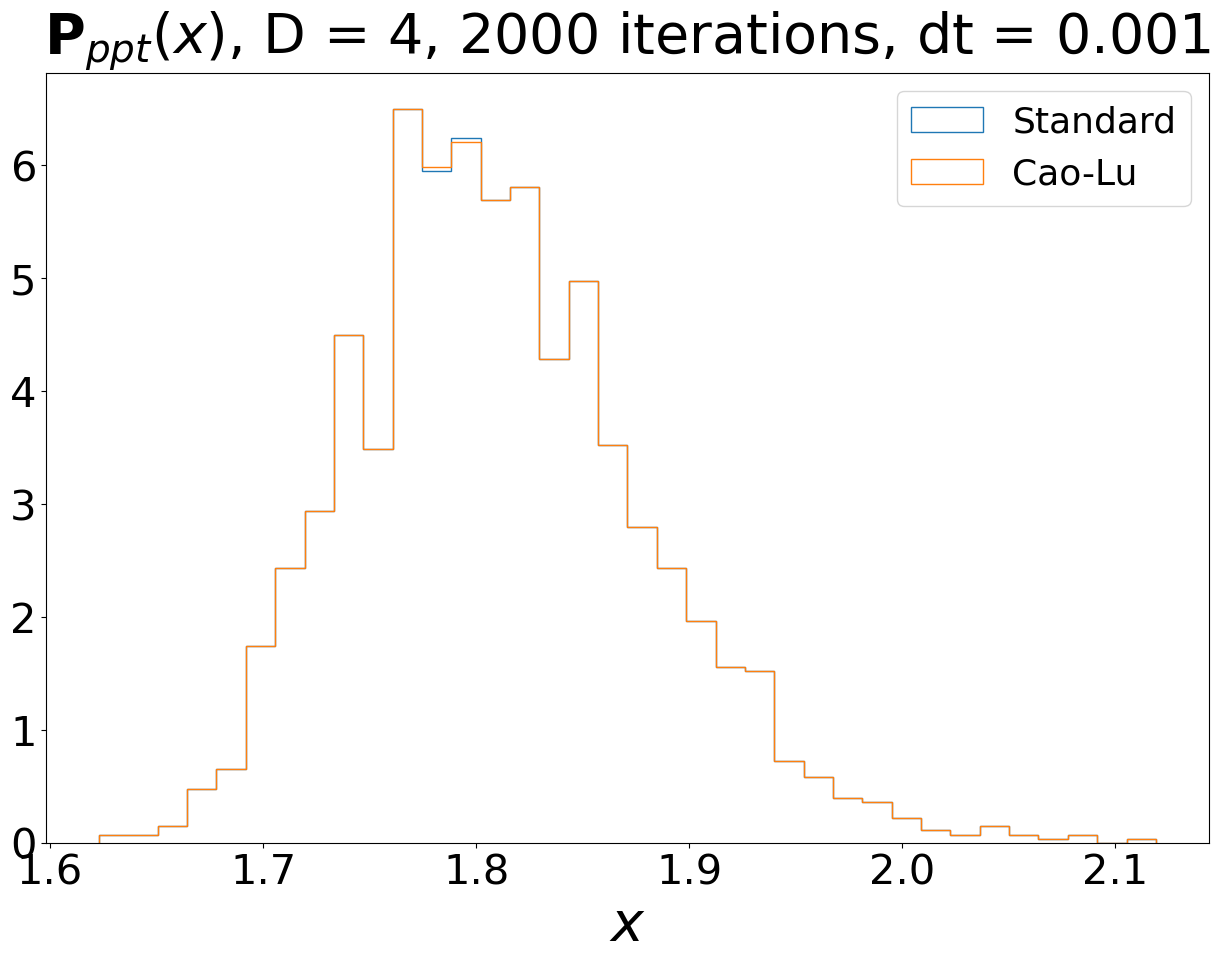}}{\label{N=4_dt=0.001}} \quad
    \subfloat[][]{\includegraphics[scale=0.225]{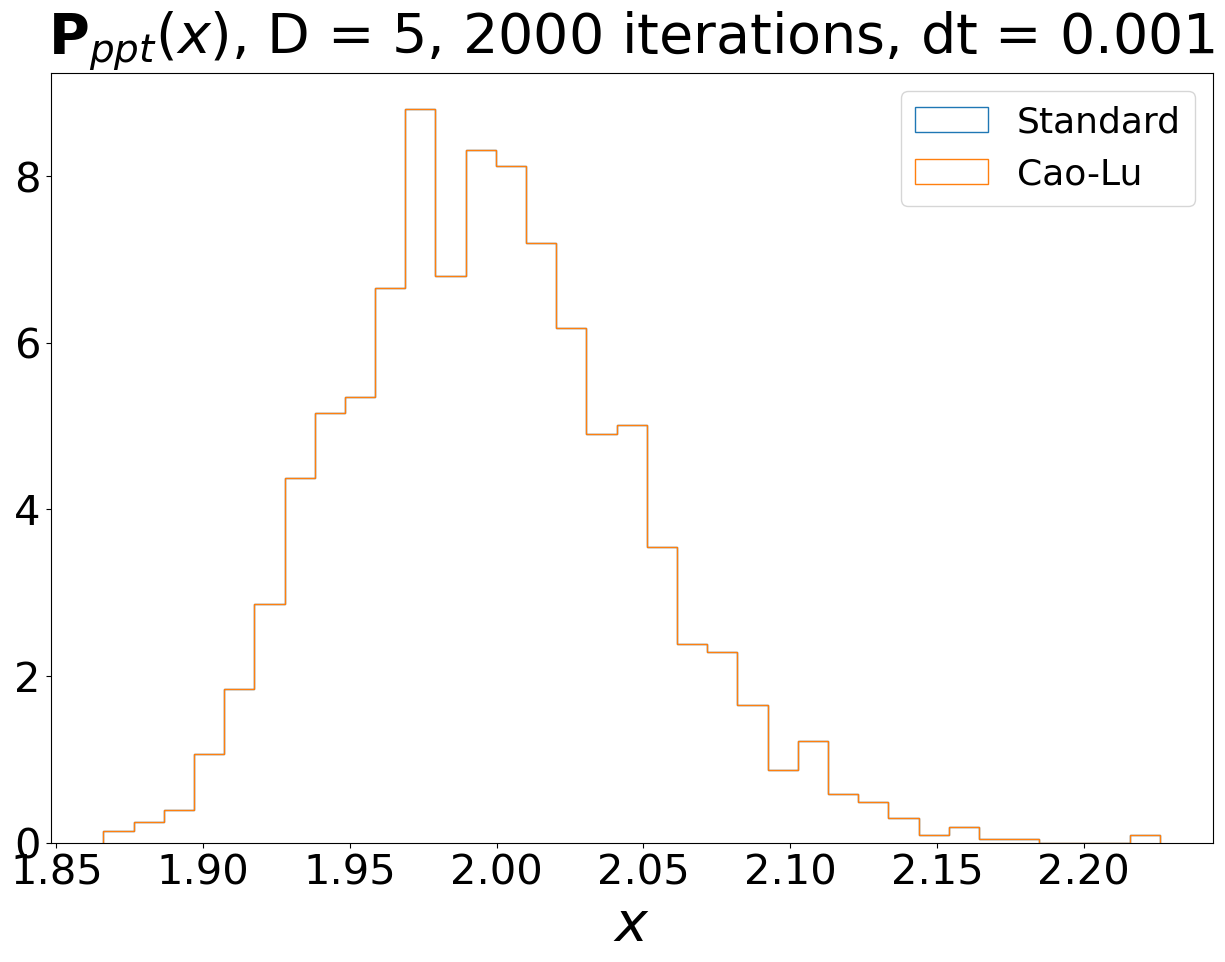}}{\label{N=5_dt=0.001}}
    \caption{Empirical rescaled PPTT distributions ${\mathbf P}_{ppt}(x)$ obtained considering $k=1$, $2000$ instances of the Lindbladian $\mathcal{L}$ and a time step $dx=0.001$.
    }
    \label{fig:EST_distrib_N=2}
\end{figure*}

This section provides numerical evidence supporting the claims of Sec.~\ref{Sect:comparison}. Specifically, we compare the empirical PPTT distributions obtained using the standard and Cao-Lu methods for different values of the subsystem dimension $D$. Our results show that when both methods are applicable, they yield the same outcomes up to numerical error. However, the Cao-Lu method is typically faster than the standard approach.
To run the simulations we set $k=1$ in Eq.~(\ref{defLL}) and
sample $\mathfrak{L}$ writing
$d\mu_D({\mathcal{L}})$ as the product of two independent measures acting respectively on the
Hamiltonian $\hat{H}$ and on the Kossakowski matrix $K$  which defines the dissipator  of the generator.
Specifically for each selected $D$, we extract  $\hat{H}$ from the Gaussian Unitary Ensemble and use
the Wishart measure $d\mu^{(\xi_D,r_D)}_D(K)$ 
for the Kossakowski matrix $K$ setting $\xi_D=D$ as scaling function in Eq.~(\ref{constraints}), and $r_D=D^2-1$ for 
Eq.~(\ref{constraints2}).

\begin{table*}
    \begin{tabular}{ccccc}
        \toprule
        Method & Parameter & Estimate & SE Estimate & 95\% CI \\
        \midrule
        Standard & $\theta_{1}$ & 0,000070 & 0,000001 & (0,000067; 0,000073) \\
        Cao-Lu & $\theta_{1}$ & 0,00095 & 0,00004 & (0,00086; 0,00105) \\
        \bottomrule
    \end{tabular}
    \caption{Parameter estimates for $T_{\text{comp}}(N)$ for the standard and the Cao-Lu method.}
    \label{tab:elapsed_comp_times_param_est}
\end{table*}

In Fig.~\ref{fig:EST_distrib_N=2} we compare the empirical rescaled PPTT distributions 
${\mathbf P}_{ppt}(x)$.
obtained with the two methods  
 for $D$ ranging from $2$ to $5$. The calculations were performed across $2000$ iterations, which represent the number of random samplings $\mathcal{L}$, using $d x=10^{-3}$ as time step increment.
From the plots it emerges that the distributions from both the standard and Cao-Lu methods are fully superimposed across all cases studied.
These results, together with previous considerations about the error in estimating $x_{ppt}:= \gamma \tau_{ppt}({\cal L})$, allow us to validate the Cao-Lu method to find the PPTTs distributions. The superimposition of the two distributions show indeed that, at least for the values of $D$ considered, a linear approximation of the negativity of entanglement near $x_{ppt}$ may be valid. 
If we suppose that such an approximation is still valid for increasing values of $D$, we can use the Cao-Lu method with the same time step of $dx=10^{-3}$, obtaining a good estimate of $x_{ppt}$ for values of $D$ up to $D=10$ (for $D=10$ we an error $\delta x_{ppt}$ of $\mathcal{O}(10^{-4})$, hence $\delta x_{ppt} \ll dx$).
However, it should be kept in mind that, for increasing values of $D$, a more complex error analysis should be done, which also takes into account the real slope of the negativity of the entanglement as a function of time. If this information is not available due to the highly non linear nature of the entanglement negativity functional, one may still choose to use a smaller $dx$ value.

\begin{figure*}
    \subfloat[][]{\includegraphics[scale=0.225]{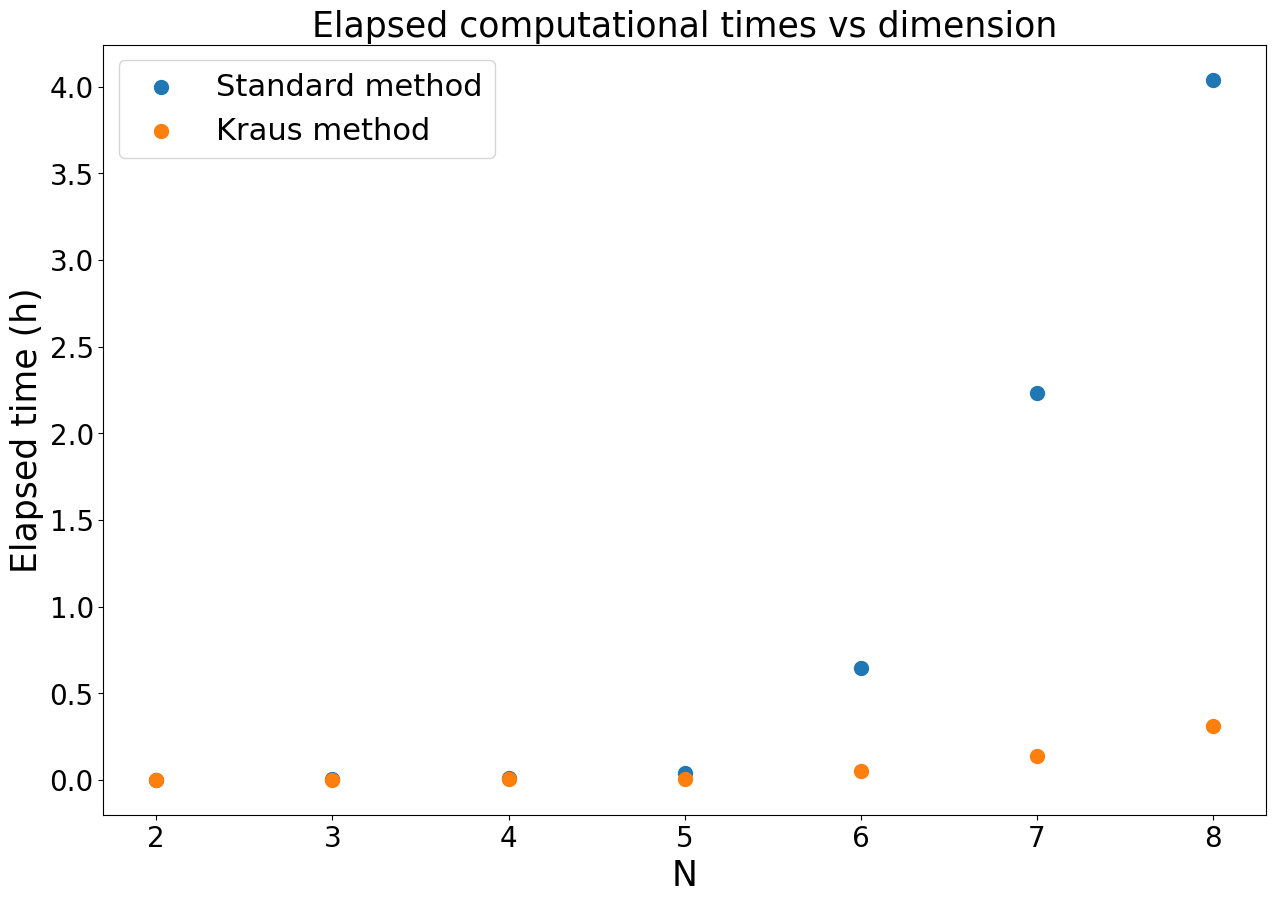}\label{fig:N_max=10_dt=0.001}} \quad
    \subfloat[][]{\includegraphics[scale=0.225]{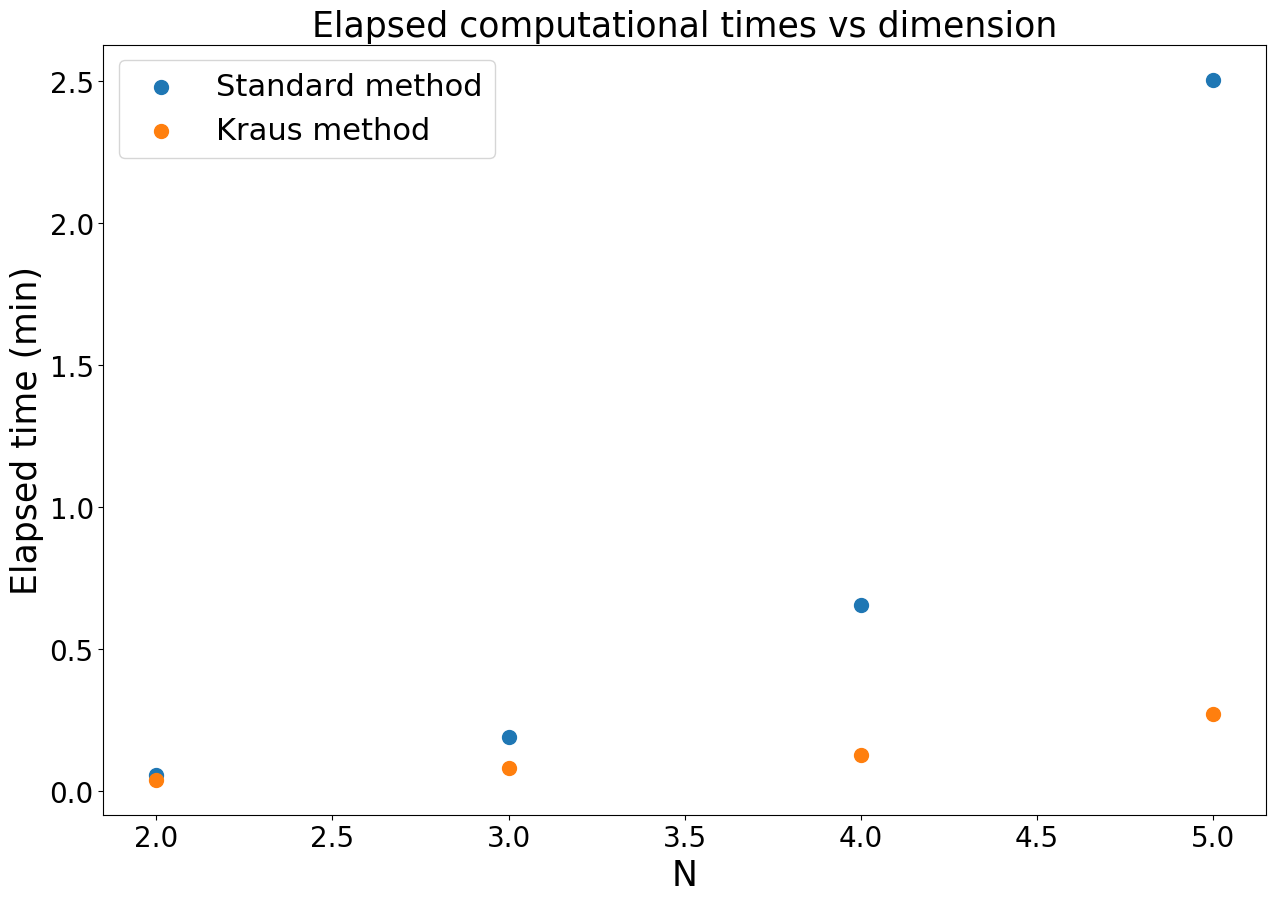}\label{fig:N_max=5_dt=0.001}}
    \caption{Elapsed computational times of the standard and the Cao-Lu method as a function of the dimension $D$ of one of the two subsystems. Number of iterations considered equal to $5$ and time step $dx=0.001$.}
    \label{fig:elapsed_times}
\end{figure*}
\begin{figure*}
    \subfloat[][]{\includegraphics[scale=0.28]{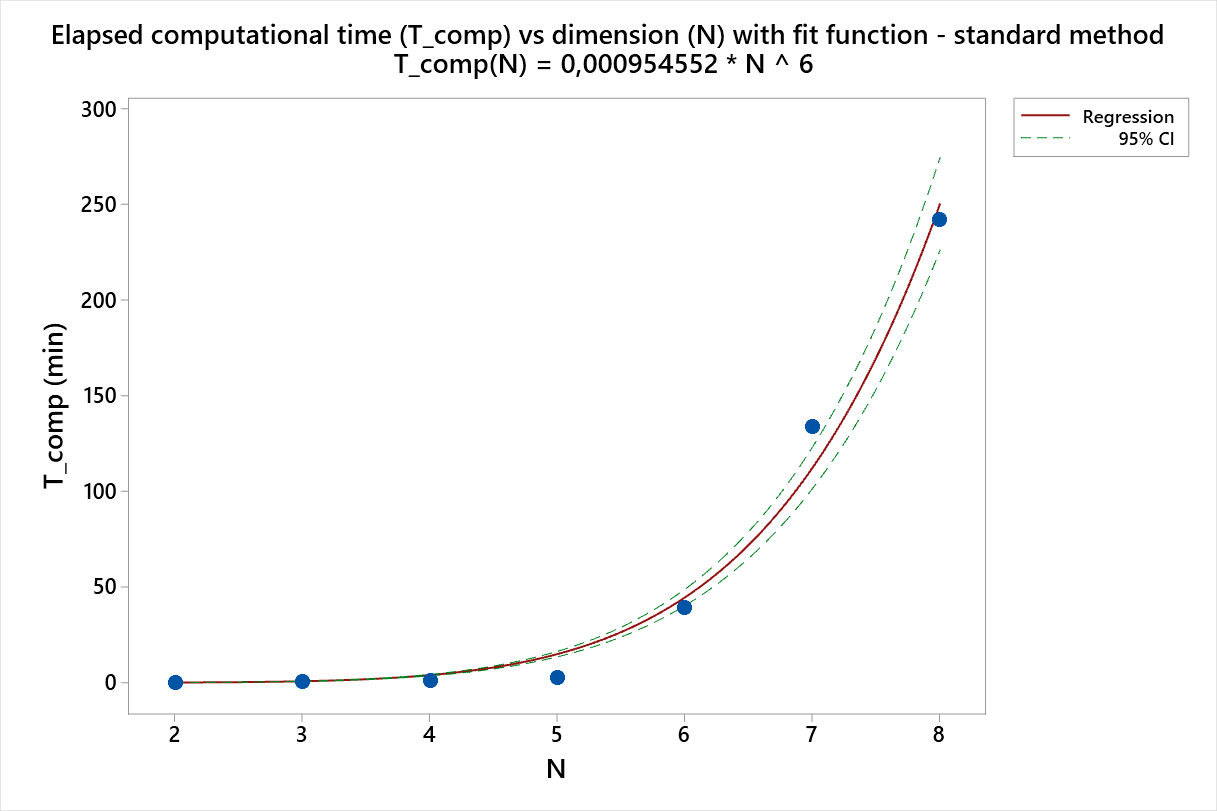}\label{fig:T_comp_fit_standard}} \quad
    \subfloat[][]{\includegraphics[scale=0.28]{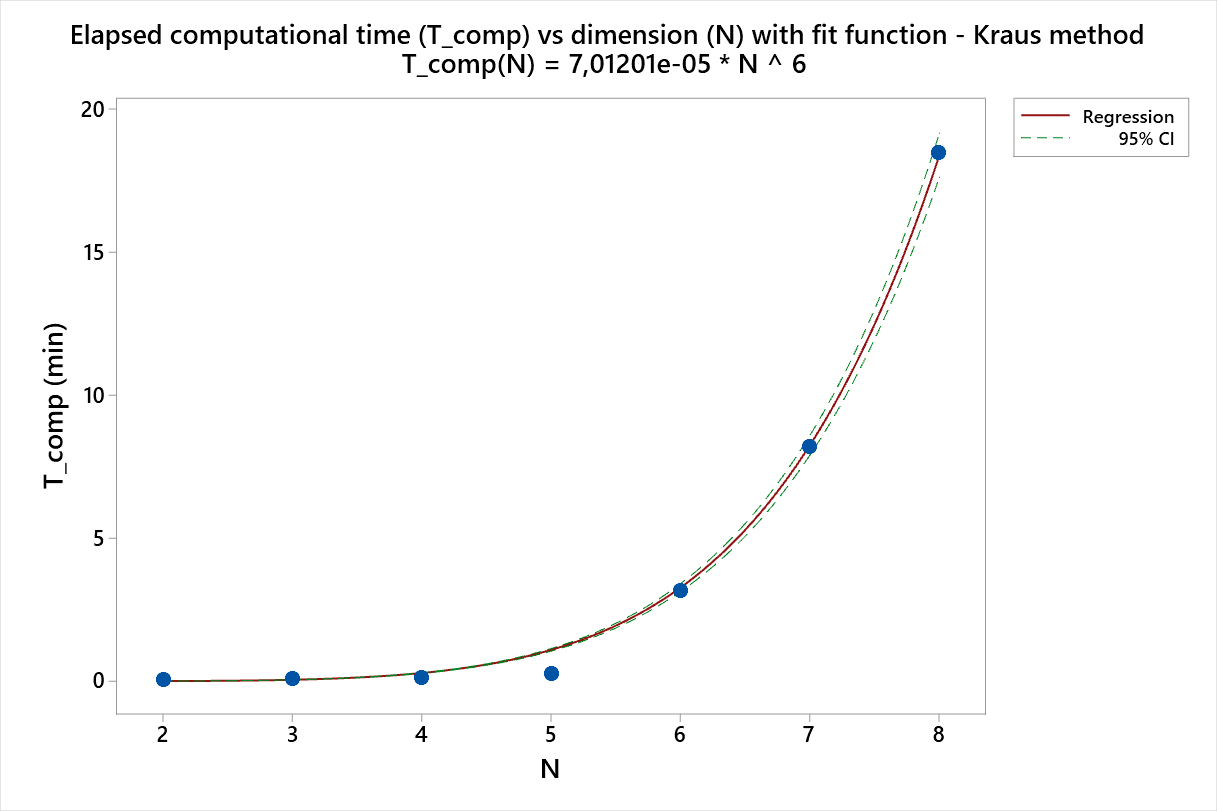}\label{fig:T_comp_fit_Kraus}}
    \caption{Elapsed computational times (in minutes) of the standard method (Panel (\ref{fig:T_comp_fit_standard})), and the Cao-Lu method (Panel (\ref{fig:T_comp_fit_Kraus})), as a function of the dimension $D$ of one of the two subsystems together with fit function. Dashed green lines represent
    the fit-curves obtained considering a 95\% CI for the fit parameters (See Table \ref{tab:elapsed_comp_times_param_est}).}
    \label{fig:elapsed_times_with_fit}
\end{figure*}

To evaluate whether the Cao-Lu method improves computational speed, we run simulations with increasing dimension 
$D$ from 2 to 8. Since the standard method's computational time grows rapidly (as evident from just a few iterations), we limit our analysis to 5 iterations -- each corresponding to a random sampling of  $\mathcal{L}$ --
using a fixed time step $dx=10^{-3}$.
Figure~\ref{fig:elapsed_times} shows the elapsed computational times $T_{\text{comp}}(D)$ for both methods. The gap between them increases with $D$, supporting the use of the Cao-Lu method for larger bipartite systems.
Given that both methods are expected to scale as $\mathcal{O}(D^{6})$  with different prefactors, we fit the computational time curves using the model $T_{\text{comp}}(D) = \theta_{1}D^{6}$. The results in Fig.~\ref{fig:elapsed_times_with_fit} 
confirm a one-order-of-magnitude difference in $\theta_{1}$,
highlighting the computational advantage of the Cao-Lu method for computing PPTT distributions.

\section{Numerical fits} 
\label{sec:fits} 

\begin{figure*}
    \subfloat[][]{\includegraphics[scale=0.305]{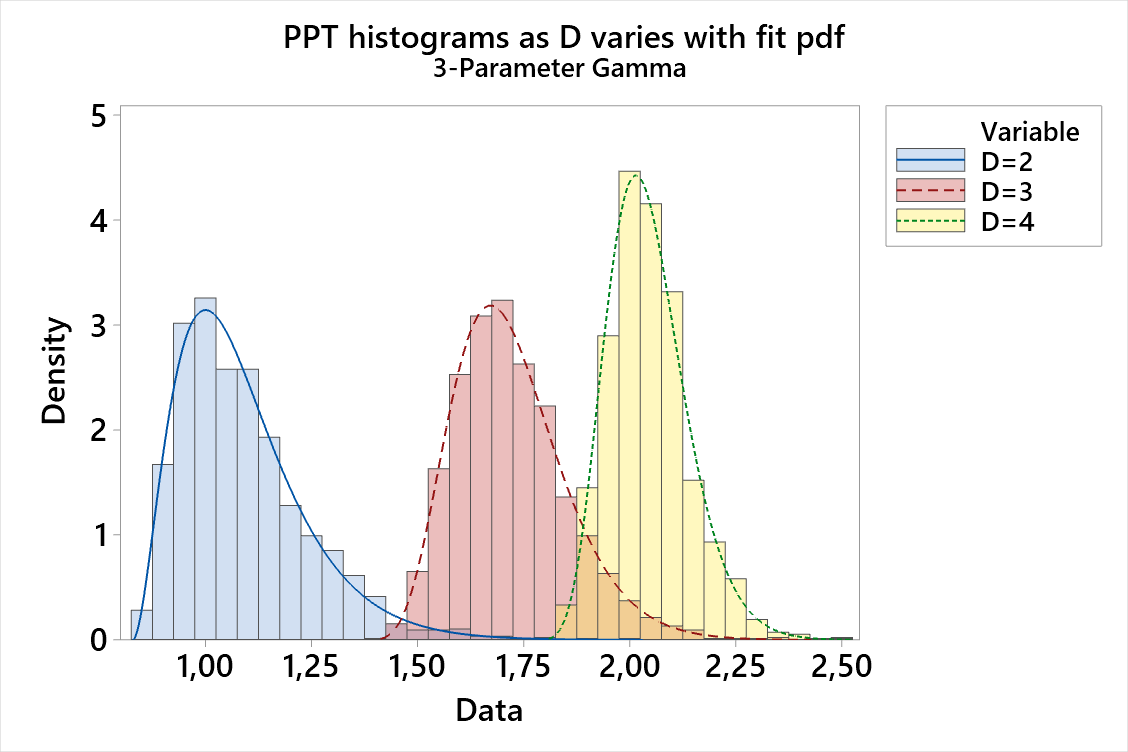}\label{fig:Hist_fit_3PGamma_N_2_4}} \quad
    \subfloat[][]{\includegraphics[scale=0.305]{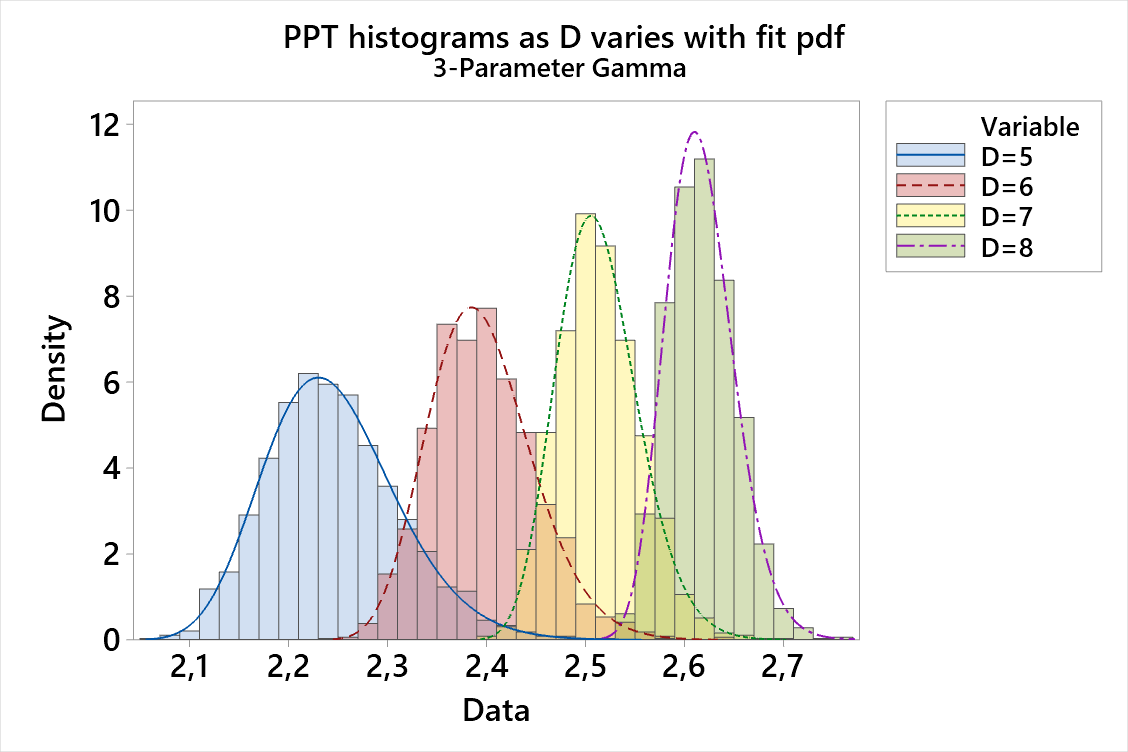}\label{fig:Hist_fit_3PGamma_N_5_8}} \\
    \subfloat[][]{\includegraphics[scale=0.305]{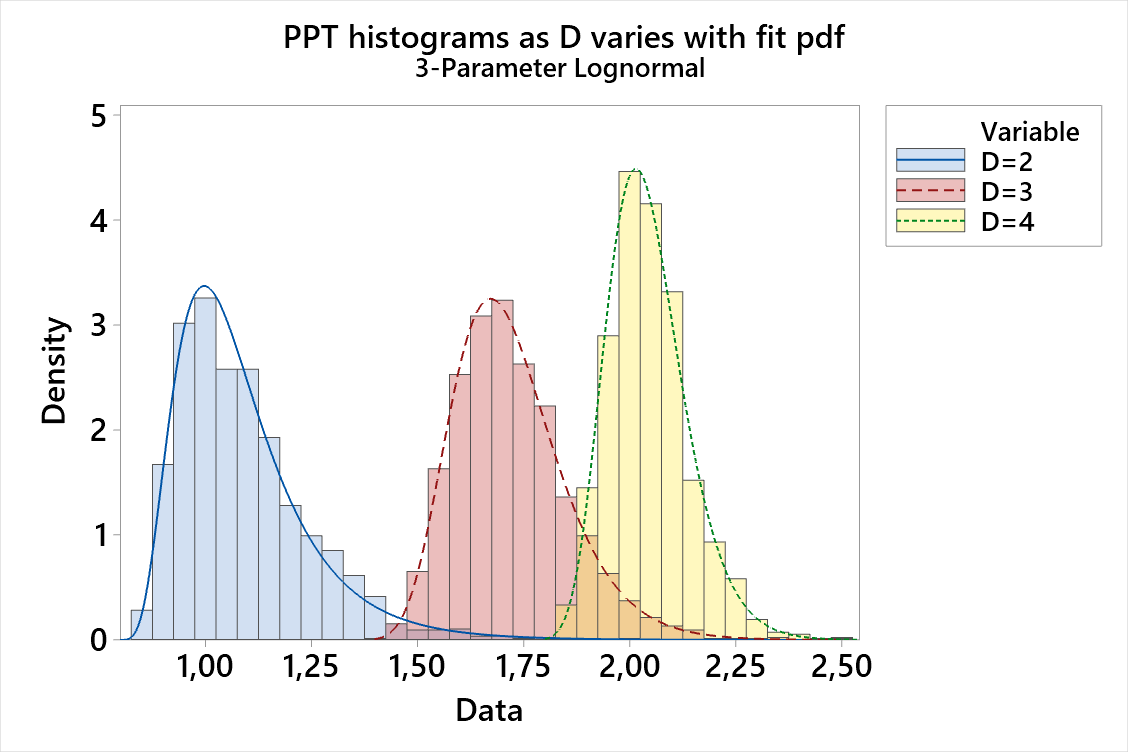}\label{fig:Hist_fit_3PLognormal_N_2_4}} \quad
    \subfloat[][]{\includegraphics[scale=0.305]{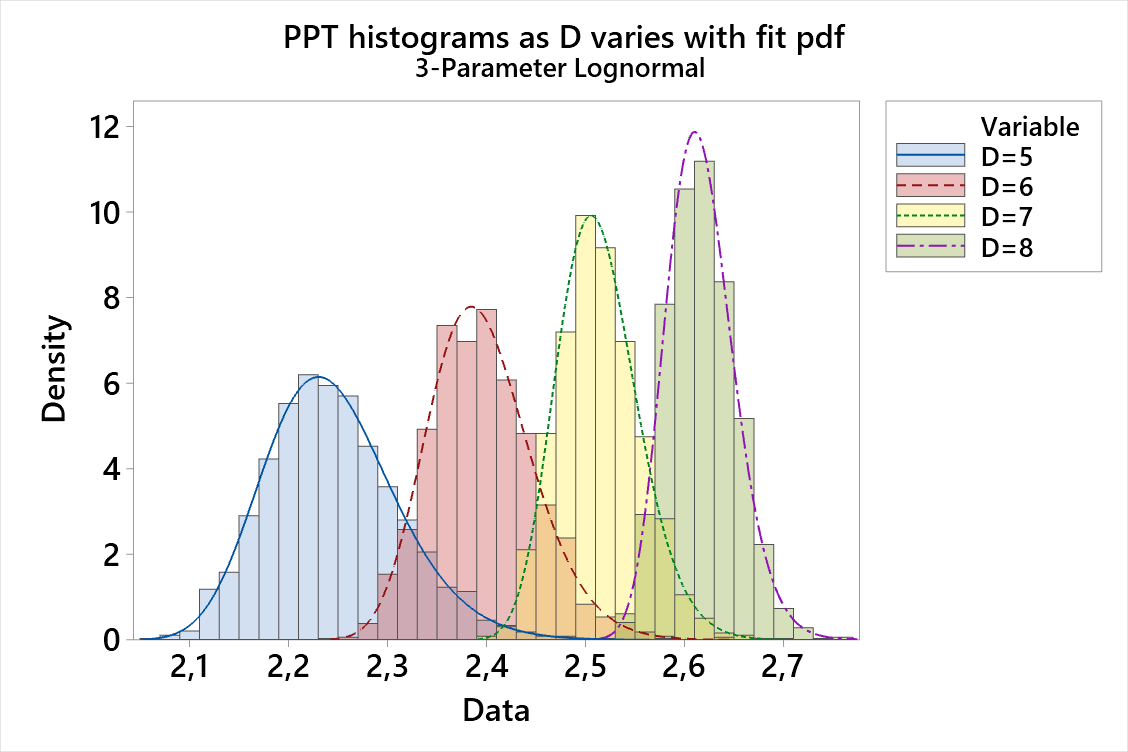}\label{fig:Hist_fit_3PLognormal_N_5_8}} 
    \caption{
    Empirical distribution ${\mathbf P}_{ppt}(x)$ for different dimensions $D$ of one subsystem with theoretical fit pdfs overlapped. Panels (\ref{fig:Hist_fit_3PGamma_N_2_4}) and (\ref{fig:Hist_fit_3PGamma_N_5_8}): The fit pdf considered is the 3-parameter Gamma distribution. Fit parameters reported in Table \ref{tab:param_estim_3PGamma}. Panels (\ref{fig:Hist_fit_3PLognormal_N_2_4}) and (\ref{fig:Hist_fit_3PLognormal_N_5_8}): The fit pdf considered is the 3-parameter Lognormal distribution. Fit parameters reported in Table \ref{tab:param_estim_3Lognormal}.} \label{fig:Histograms_as_N_varies_with_fit_pdfs}
\end{figure*}

In this section we present a numerical fit of the 
 empirical distributions reported in Sec.~\ref{Sect:PPTT_RandomNoise}.
 To carry on this analysis, we use the Minitab software \cite{minitab}. This software enables the testing of up to 14 theoretical distributions, employing the maximum likelihood (ML) method to determine the optimal parameters of the selected model. Additionally, probability plots and goodness-of-fit tests are employed to ascertain the distribution that best fits the data.
From this analysis it emerges that the two distributions that could represent a good fit for the data are the 3-parameter Gamma distribution (Eq.~\eqref{Eq:3P-Gamma}) and the 3-parameter Lognormal distribution (Eq.~\eqref{Eq:3P-Lognormal}), with no elements that justify the propensity for one distribution rather than the other. We report in the following the functional form of such probability distribution functions.
\begin{equation}\label{Eq:3P-Gamma}
    f_{G}^{(\beta,\sigma,\mu)}(x) := \frac{(x-\mu)^{(\beta-1)}}{\sigma^{\beta}\Gamma(\beta)}\exp\Big(-\frac{(x-\mu)}{\sigma}\Big),
\end{equation}
\begin{eqnarray}\label{Eq:3P-Lognormal}
    f_{LN}^{(\sigma,\mu,\nu)}(x) &:=& \frac{(x-\mu)^{-1}}{\sqrt{2\pi} \sigma}\\ \nonumber
    &&\times \exp\Big(-\frac{(\ln(x-\mu) - \nu)^2}{2\sigma^2}\Big).
\end{eqnarray}
Here $\beta$ is the shape parameter, $\sigma$ is the scale parameter, $\mu$ is the threshold value and $\nu$ is the location. In reaching this conclusion, the statistical analysis involves generating a probability plot for each tested distribution, associated to which Minitab calculates the Anderson-Darling coefficient (AD) and the corresponding p-value. The AD coefficient measures the discrepancy between the empirical and theoretical distributions, computed based on cumulative distribution. Smaller AD values indicate better alignment between the data and the theoretical distribution. Hence, among the fit distributions analyzed, those with lower AD values are considered better than others.
We report in Fig.~\ref{fig:Histograms_as_N_varies_with_fit_pdfs}  the plots that we obtain when considering the empirical PPTT distributions ${\mathbf P}_{ppt}(x)$ together with the fit pdfs overlapped.
\begin{table}[H]
    \centering
    \begin{tabular}{cccc}
        \toprule
        D & Shape $(\beta)$ & Scale $(\sigma)$ & Threshold $(\mu)$ \\
        \midrule
        2 & 2,240 & 0,06972 & 0,8334\\
        3 & 7,122 & 0,03602 & 1,280\\
        4 & 17,32 & 0,01644 & 1,526\\
        5 & 20,86 & 0,01099 & 1,769\\
        6 & 12,53 & 0,01150 & 2,000\\
        7 & 18,03 & 0,007438 & 2,132\\
        8 & 27,74 & 0,004910 & 2,237\\
        \bottomrule
    \end{tabular}
    \caption{Maximum likelihood estimates of the 3-parameter Gamma distribution for ${\mathbf P}_{ppt}(x)$ at different subsystem dimensions $D$.}
    \label{tab:param_estim_3PGamma}
\end{table}

\begin{table}[H]
    \centering
    \begin{tabular}{cccc}
        \toprule
        D & Location $(\nu)$ & Scale $(\sigma)$ & Threshold $(\mu)$\\
        \midrule
        2 & -1,899 & 0,5856 & 0,8120 \\
        3 & -1,088 & 0,2718 & 1,186 \\
        4 & -0,9229 & 0,1688 & 1,408 \\
        5 & -1,093 & 0,1474 & 1,659 \\
        6 & -1,573 & 0,1910 & 1,933 \\
        7 & -1,669 & 0,1645 & 2,075 \\
        8 & -1,638 & 0,1314 & 2,177 \\
        \bottomrule
    \end{tabular}
    \caption{Maximum likelihood estimates of the 3-parameter Lognormal distribution for ${\mathbf P}_{ppt}(x)$ at different subsystem dimensions $D$.}
    \label{tab:param_estim_3Lognormal}
\end{table}

\section{Normalization analysis}\label{sec:norm} 

In this section we give explicit proof of the identity~(\ref{block}) and show that 
choosing the super-linear scaling  ${\rm x} \log_2 {\rm x}$
for the normalization constraint~(\ref{constraints})
 ensures that the Kossakowski matrices associated with the PPTT distributions in the iLOC, cLOC, and GBL scenarios -- introduced in Sec.~\ref{sec:LNvsGN} -- all satisfy the same normalization condition. 
The starting point of the analysis is to relate the $(D^2-1)\times (D^2-1)$ Kossakowski matrix $K^{[\rm{LOC}]}$ of the generator ${\cal D}^{[\rm{LOC}]}_{K_1,\cdots, K_n}$ in Eq.~(\ref{locL}) to the $(d_j^2-1) \times (d_j^2-1)$ matrices $K_j$ of its local constituents ${\cal D}_{K_j}^{(j)}$.
To achieve this, we find it useful to identify 
the orthonormal set of traceless operators, $\{\hat{F}_{m}\}_{m=1,\cdots,D^2 -1}$ of $A$
which enters in Eq.~(\ref{eq:Dissipator_Kmn}) as a tensor product of 
 local terms $\{\hat{F}^{(j)}_{\ell}\}_{\ell=0,\cdots,d_j^2 -1}$ formed by the local complete orthonormal set
of the traceless operators we used  to define the $K_j$, plus the extra element 
 \begin{eqnarray}
\hat{F}^{(j)}_{0} = \hat{\mathbb{1}}^{(j)}/\sqrt{d_j} \;,
\end{eqnarray} 
where $\hat{\mathbb{1}}^{(j)}$ is the identity operator on subsystem $a_j$. 
Specifically we write  the $\hat{F}_{m}$'s as 
\begin{eqnarray}\label{product} 
\hat{F}_{m} := \hat{F}^{(1)}_{\ell_1} \otimes \hat{F}^{(2)}_{\ell_2} \otimes \cdots \otimes \hat{F}^{(n)}_{\ell_n}\;,
\end{eqnarray} 
identifying the label $m$ with the $n$-dimensional vector $(\ell_1,\ell_2,\cdots, \ell_n)$. Observe 
 that the operator~(\ref{product}) is
traceless iff at least one of the components $\ell_j$ is different from $0$, and that 
for generic $m=(\ell_1,\ell_2,\cdots, \ell_n)$ and $m'=(\ell'_1,\ell'_2,\cdots, \ell'_n)$ we have indeed that
\begin{eqnarray}\label{product2} 
\mbox{Tr}[\hat{F}^\dag_{m} \hat{F}_{m'}] &=& \prod_{j=1}^n \mbox{Tr}[\hat{F}^{(j)\dag}_{\ell_j}
\hat{F}^{(j)\dag}_{\ell'_j}]  \\&=& \prod_{j=1}^n\delta_{\ell_j,\ell'_j} = \delta_{m,m'}\;. \nonumber 
\end{eqnarray} 
The next step is to observe that the local dissipator acting on the $j$-th subsystem behaves as the identity superoperator on all other subsystems. Thus, denoting the matrix elements of $K_j$ as $[K_j]_{\ell,\ell'}$, we can write: 
    \begin{eqnarray}\nonumber
        &&\!\!{\cal D}^{(j)}_{K_j}(\,\cdots\,) =  \sum_{\ell,\ell'=0}^{d_j^2 -1}  [K_j]_{\ell,\ell'} \Big( \hat{\mathbf F}^{[j]}_{\ell'}(\,\cdots\,)\hat{\mathbf F}^{[j]\dagger}_{\ell} \\
        &&\!\! - \frac{1}{2}\left( \hat{\mathbf F}^{[j]\dagger}_{\ell}\hat{\mathbf F}^{[j]}_{\ell'}(\,\cdots\,) +(\,\cdots\,) 
        \hat{\mathbf F}^{[j]\dagger}_{\ell}\hat{\mathbf F}^{[j]}_{\ell'} \right) \Big). \nonumber   \\\label{eq:Dissipator_Kmnj}
    \end{eqnarray}
Here $\hat{\mathbf F}^{[j]}_{\ell}$ 
is the operator acting as the identity on all subsystems except  $a_j$, where it acts as $\hat{F}^{(j)}_{\ell}$: 
\begin{eqnarray}\nonumber 
&&\hat{\mathbf F}^{[j]}_{\ell} := \hat{\mathbb{1}}^{(1)}\otimes \hat{\mathbb{1}}^{(2)}\otimes \cdots \otimes  \hat{F}^{(j)}_{\ell}  \otimes \cdots
\otimes \hat{\mathbb{1}}^{(n)} \\
&&= \sqrt{\tfrac{{D}}{{d_j}}} \hat{F}^{(1)}_0\otimes  \hat{F}^{(2)}_0\otimes \cdots \otimes  \hat{F}^{(j)}_{\ell}  \otimes \cdots
\otimes  \hat{F}^{(n)}_0\;, \nonumber \\\label{ide11} 
\end{eqnarray} 
where we used the fact that 
\begin{eqnarray}
D=\prod_{j'=1}^n d_{j'}\;. \label{dimension} 
\end{eqnarray} 
Replacing~(\ref{eq:Dissipator_Kmnj}) into Eq.~(\ref{locL})  we can hence write
 \begin{eqnarray}\nonumber 
&&\!\! \!\! \!\! {\cal D}^{[\rm{LOC}]}_{K_1,\cdots, K_n}(\cdots) = \sum_{j=1}^n 
\sum_{\ell,\ell'=0}^{d_j^2 -1}  [K_j]_{\ell,\ell'} \Big( \hat{\mathbf F}^{[j]}_{\ell'}(\,\cdots\,)\hat{\mathbf F}^{[j]\dagger}_{\ell} \\
        &&\!\!   - \frac{1}{2}\left( \hat{\mathbf F}^{[j]\dagger}_{\ell}\hat{\mathbf F}^{[j]}_{\ell'}(\,\cdots\,) +(\,\cdots\,) 
        \hat{\mathbf F}^{[j]\dagger}_{\ell}\hat{\mathbf F}^{[j]}_{\ell'} \right) \Big),\nonumber   \\
        \label{eq:Dissipator_Kmnjasd}
\end{eqnarray} 
which can be casted in the form~(\ref{eq:Dissipator_Kmn}) with a  Kossakowski matrix 
${K}^{[\rm{LOC}]}$ with elements 
\begin{eqnarray}{K}^{[\rm{LOC}]}_{m,m'} := \left\{ \begin{array}{ll} \label{defKLOC} 
 \frac{D}{d_j} [K_j]_{\ell,\ell'}&  \forall m,m' \in {\cal S}_j, \\ \\0 & \mbox{otherwise,}  
 \end{array} \right.
\end{eqnarray} 
with ${\cal S}_j$ the set identified by the vectors $(\ell_1,\ell_2,\cdots, \ell_n)$ that can have non-zeros values only on the $j$-th
entry. 
Since for $j\neq j'$,   ${\cal S}_j$ and ${\cal S}_{j'}$ have zero-overlap, the above equation implies that 
 ${K}^{[\rm{LOC}]}$ has the block structure anticipated in Eq.~(\ref{block}) of the main text ensuring the validity of the relations
 ~(\ref{block11}). 
For instance a direct proof of the first of such expressions can be obtained from~(\ref{defKLOC}) by observing that 
\begin{eqnarray}\label{asdf} 
\!\!\mbox{Tr}[{K}^{[\rm{LOC}]}] &=& \sum_{j=1}^n \sum_{m\in {\cal S}_j} {K}^{[\rm{LOC}]}_{m,m} \\
&=&\sum_{j=1}^n \sum_{\ell}  \frac{D}{d_j} [K_j]_{\ell,\ell}
=\sum_{j=1}^n  \frac{D}{d_j}  \mbox{Tr}[K_j] \;. \nonumber
\end{eqnarray} 
Therefore if we sample the $K_j$'s under the normalization condition~(\ref{xiJ}) in iLOC scenario
thanks to (\ref{dimension}) we get that ${K}^{[\rm{LOC}]}$ has the same
trace of the Kossakowski matrices we sample in the GBL scenario, i.e. 
\begin{eqnarray}\label{final} 
\mbox{Tr}[{K}^{[\rm{LOC}]}] &=&  \sum_{j=1}^n  \frac{D}{d_j}  \xi_{d_j} = D \sum_{j=1}^n\log_2 d_j \nonumber \\
 &=& D \log_2 D \;.
\end{eqnarray} 
The same conclusion applies also in the cLOC scenario, simply here all the subsystems have the same dimension $d$, and we act on them with the same
local Kossakowski matrix $K_1$, so that once~(\ref{xiJ}) is enforced Eq.~(\ref{asdf}) leads once more to (\ref{final}).

\end{document}